\documentclass[pre,aps,twocolumn,showpacs]{revtex4}
\usepackage{epsfig}
\usepackage{graphicx}

\begin{document}
\title{Parametric Autoresonance in Faraday waves}
\author{Michael Assaf and Baruch Meerson}
\affiliation{Racah Institute of Physics, Hebrew University of
Jerusalem, Jerusalem 91904, Israel} 
\begin{abstract}
We develop a theory of parametric excitation of weakly nonlinear
standing gravity waves in a tank, which is under vertical
vibrations with a slowly time-dependent (``chirped") vibration
frequency. We show that, by using a negative chirp, one can excite
a steadily growing wave via parametric autoresonance. The method
of averaging is employed to derive the governing equations for the
primary mode. These equations are solved analytically and
numerically, for typical initial conditions, for both inviscid and
weakly viscous fluids. It is shown that, when passing through
resonance, capture into resonance always occurs when the chirp
rate is sufficiently small. The critical chirp rate, above which
breakdown of autoresonance occurs, is found for different initial
conditions. The autoresonance excitation is expected to terminate
at large amplitudes, when the underlying constant-frequency system
ceases to exhibit a non-trivial stable fixed point. 
\end{abstract}
 \maketitle

\section{Introduction} There are two ways of driving
a classical nonlinear oscillator by a small oscillating force: via
either external or parametric resonance. In both cases, the
initial growth of the amplitude of the oscillator is arrested,
even without dissipation, when nonlinear effects come into play.
This is due to the fact that the natural frequency of a nonlinear
oscillator is amplitude-dependent, so a mismatch between the
(invariable) driving frequency and the natural frequency appears
\cite{Landau1}.

To overcome the nonlinear mismatch and maintain phase locking
between the driving force and the oscillator, one can slowly vary
the driving frequency with time so as to achieve a persistent
growth of the oscillations. This simple and versatile method is
called autoresonance: either external, or parametric. To emphasize
the difference between the two, let us watch a child on a swing.
When a parent pushes the swing (once in each cycle), he gradually
increases the time interval between the pushes as the swing
amplitude grows. Here he employs external autoresonance. On the
contrary, when the child swings himself (he achieves it by moving
the position of his center of mass up and down \textit{twice} in
each cycle), he gradually increases the period of these
modulations as the swing amplitude grows. This is parametric
autoresonance.

The simple model of a nonlinear oscillator, excited via external
autoresonance, has found numerous applications in physics, see
\textit{e.g.} Ref. \cite{Friedland1} for a brief review. The
external autoresonance scheme has been also generalized to systems
with an infinite number of degrees of freedom, such as nonlinear
waves \cite{Meerson5,Meerson3,Lazar_3wave} and vortices
\cite{vortices}. In contrast to the external autoresonance,
parametric autoresonance has received much less attention
\cite{Khain}. In this work we generalize to nonlinear waves the
theory, developed in Ref. \cite{Khain} for a nonlinear oscillator.
Specifically, we show that the parametric autoresonance mechanism
can be used for driving nonlinear standing gravity waves with a
steadily growing amplitude on a free surface of a fluid.

Michael Faraday \cite{Faraday} was the first to observe that, when
a tank containing a fluid is periodically vibrated in the vertical
direction, a standing wave pattern forms at the free surface of
the fluid when the vibration frequency is twice the frequency of
the surface vibrations. This phenomenon is a classic example of
parametric resonance, because the vertical acceleration of the
tank - an intrinsic parameter of the system - depends on time via
the periodic vibration. Lord Rayleigh \cite{Rayleigh} carried out
a further series of experiments, which supported Faraday's
observations, and also developed a linear theory for these waves
in terms of a linear Mathieu equation. Benjamin and Ursell
\cite{Ursell} advanced the linear theory further. Subsequently,
Miles \cite{Miles1,Miles2,Miles3}, Douady \cite{douady}, Milner
\cite{milner}, and Decent and Craik \cite{decent} formulated a
weakly nonlinear theory based on amplitude expansion, while some
of these and indeed numerous other works dealt with experimental
studies of Faraday waves.

Being interested in parametric \textit{autoresonance}, we add a
new dimension to the problem of Faraday waves and investigate
weakly nonlinear standing gravity waves formed when the vibration
frequency is slowly \textit{decreased} (chirped downwards) in
time. We show that the negative frequency chirp causes a
persistent growth of the wave amplitude. Like in other instances
of autoresonance, the exact form of the frequency chirp is
unimportant once the chirp sign is correct, and the chirp rate is
not too high. The autoresonance excitation is expected to
terminate at large amplitudes, when an underlying dynamical
system, corresponding to the case of a  \textit{constant}
frequency, ceases to exhibit a non-trivial stable fixed point.

Here is the layout of the rest of the paper. Section II presents a brief overview of theory of weakly nonlinear
Faraday waves with a constant driving frequency. Sections III and IV deal with theory of chirped Faraday waves,
in inviscid (Sec. III) and low-viscosity (Sec. IV) fluids. Section V presents a brief discussion of our results.
\vspace{-0.5cm}
\section{Weakly nonlinear Faraday waves}
\vspace{-0.2cm}
\subsection{Inviscid fluid}
\vspace{-0.2cm} To set the stage for a theory of chirped Faraday waves, we need to briefly review the theory of
weakly nonlinear Faraday waves with a constant driving frequency. Consider a quasi-two-dimensional rectangular
tank with a fluid of length $l$, width $w$ and depth $h$, so that $l \gg w$. We assume that the elevation of the
fluid, caused by the wave, depends only on the longitudinal coordinate $x$ and time $t$, so that we have a
quasi-two-dimensional flow in the $xz$ plane ($z$ is the vertical coordinate). The unperturbed level of the
fluid is at $z=0$. The vertical displacement of the vibrating tank is described by the equation
\begin{equation}\label{zeta}
\zeta(t)=a_{0}\cos(2\omega\,t)\,.
\end{equation}
We assume weak forcing, that is the vibration acceleration is much
less than the gravity acceleration $g$, and introduce a small
dimensionless parameter $\varepsilon$:
\begin{equation}\label{eps}
\varepsilon=\frac{\omega^{2}a_{0}}{g}\ll\,1\,.
\end{equation}
In the limit of inviscid fluid the flow remains potential once it
is potential at $t=0$, and the external forces are potential
\cite{Lamb}, as is the case here. The assumption of a potential
flow is also approximately valid in a low-viscosity fluid
\cite{Ursell}. We also assume that the wavelength of the standing
wave is much larger than the capillary length of the fluid and
neglect the capillary effects throughout the paper. The linear
dispersion relation for the wave is $\omega_{n}^{2}=\,gk_{n} \tanh
(k_n h)$, where $\omega_{n}$ is the natural frequency of the
$n$-th mode, $k_{n}= 2\pi/\lambda_{n}=n\pi/l$ is the wave number
of the $n$-th mode, and $n=1,2,\dots$.

The governing equations for the velocity potential
$\varphi(x,z,t)$ and the wave profile $\eta(x,t)$ are
\cite{Ursell,Lamb,Landau2,Currie}:
\begin{eqnarray}
\nabla^{2}\varphi&=&0\,, \label{nonlin1}\\
\left.\left[\varphi_{t}+\frac{1}{2}(\varphi_{x}^{2}+\varphi_{z}^{2})+(g+\ddot{\zeta})\eta\right]\right|_{_{z=\eta}}&=&0\,, \label{nonlin2}\\
\left(\eta_{t}+\varphi_{x}\eta_{x}-\varphi_{z}\right)|_{_{z=\eta}}&=&0 \,,\label{nonlin3}\\
\varphi_{z}\,|_{_{z=-h}}&=&0\,,\label{nonlin4}
\end{eqnarray}
where indices denote partial derivatives. The Laplace's equation
(\ref{nonlin1}) describes a potential flow of an incompressible
fluid. Equations (\ref{nonlin2}) and (\ref{nonlin3}) are the
Navier-Stokes equation and the kinematic boundary condition,
respectively, evaluated at the free surface. Finally, Eq.
(\ref{nonlin4}) is the boundary condition for the vertical
velocity component at the bottom of the tank.

Let the vibration frequency be close to twice the natural
frequency of the primary mode $n=1$, \textit{i.e.}
$\omega\simeq\omega_{1}$, so that this mode is excited via
parametric resonance. In a weakly nonlinear regime it suffices to
account for the excitation of only one higher order mode: the
secondary mode $n=2$, which is enslaved to the primary mode
\cite{Miles2}. Therefore, one should look for $\varphi(x,z,t)$ and
$\eta(x,t)$ in Eqs. (\ref{nonlin1})-(\ref{nonlin4}) in the
following form \cite{Miles1,Miles2}:
\begin{eqnarray}\label{waveprof}
\varphi(x,z,t)\!&=&\!\varphi_{0}(t)\!+\!\varphi_{1}(t)\psi_{1}(x)\frac{\cosh\left[k_{1}(z\!+\!h)\right]}{\cosh(k_{1}h)}+\nonumber\\
&+&\varphi_{2}(t)\psi_{2}(x)\frac{\cosh\left[k_{2}(z\!+\!h)\right]}{\cosh(k_{2}h)} + \dots \,,\nonumber\\
\eta(x,t)\!&=&\!\eta_{1}(t)\psi_{1}(x)+\eta_{2}(t)\psi_{2}(x)+
\dots \,,
\end{eqnarray}
where the eigenfunctions $\psi_{n}(x)=\sqrt{2}\,\cos\,(k_{n}x)$.
The higher order terms will be neglected in the following. In the
deep-water limit $k_n h \gg 1$, the linear dispersion relation for
the wave becomes $\omega_{n}^{2}\simeq \,gk_{n}$. For this
approximation to hold with an error less than 0.5\%, it suffices
to demand that $h>l$.

The perturbation theory we are using employs the smallness of
$\varepsilon$. As will be seen later, this smallness implies a
smallness of the wave amplitude compared with the wavelength, so
that the dimensionless parameter $\kappa = k_{1}\eta$ is small.
Expanding $\varphi(x,z=\eta,t)$ in the vicinity of the unperturbed
surface $z=0$ in a power series in $\kappa$, and substituting it
and the second of Eqs. (\ref{waveprof}) into Eqs. (\ref{nonlin2})
and (\ref{nonlin3}), we obtain in the leading and sub-leading
orders of $\kappa$:
\begin{equation}\label{enslaving}
\eta_{2}(t)\simeq \frac{k_{1}\eta_{1}^{2}(t)}{\sqrt{2}}\,,
\end{equation}
\begin{equation}
\varphi_{0}(t) \simeq
-\dot{\eta}_{1}\eta_{1}\,,\;\;\varphi_{1}(t)\simeq\frac{\dot{\eta}_{1}}{k_{1}}\,,\;\;\varphi_{2}(t)\simeq
0\,. \label{relations}
\end{equation}
As we see, the next-order corrections $\varphi_{0}$ and $\eta_{2}$
are enslaved to the primary mode, and their magnitudes are ${\cal
O}(\kappa \eta_{1})$. In addition, we obtain a nonlinear
differential equation of the second-order for the time-dependent
amplitude of the primary mode $\eta_{1}(t)$:
\begin{equation}\label{motion}
\ddot{\eta}_{1}+\frac{1}{2}k_{1}^{2}(5\dot{\eta}_{1}^{2}\eta_{1}-3\omega_{1}^{2}\eta_{1}^{3})+
\omega_{1}^{2}\left[1+4\varepsilon\cos(2\omega\,t)\right]\eta_{1}=0\,,
\end{equation}
where we have used Eq. (\ref{zeta}) and kept terms up to ${\cal O}(\kappa^{3})$. Equation (\ref{motion}) is a
generalization of the linear Mathieu equation \cite{Bogoliubov}.

Now we employ the method of averaging \cite{Bogoliubov}. We make
an Ansatz
$\eta_{1}(t)=A_{1}(t)\cos\left[\omega_{1}t+\phi_{1}(t)\right]$ and
$\dot{\eta}_{1}(t)=-\omega_{1}A_{1}(t)\sin\left[\omega_{1}t+\phi_{1}(t)\right]$
in Eq. (\ref{motion}) and treat the amplitude $A_{1}(t)$ and phase
$\phi_{1}(t) $ as slow functions of time. (Being interested in the
first-order equations with respect to $\varepsilon$, one can omit
higher temporal harmonics in $\eta_{1}(t)$
\cite{Bogoliubov,perturbative}.) Let
$\delta\!=\!\omega_{1}\!-\!\omega\,$ be the (small) detuning from
the exact linear resonance. Then, for $\varepsilon\ll\,1$ and
$|\delta|\ll\omega_{1}$, we can perform averaging over the fast
time $\sim \omega_1^{-1}$ \cite{Bogoliubov}. Introducing a new
phase variable $\phi=\delta\cdot\,\!t+\phi_{1}$, we obtain:
\begin{eqnarray}
\dot{A_{1}}&=&\varepsilon\omega_{1}A_{1}\sin(2\phi)\,,\nonumber\\
\dot{\phi}&=&\varepsilon\omega_{1}\cos(2\phi)-\frac{k_{1}^{2}A_{1}^{2}\omega_{1}}{4}+\delta\,. \label{averaged1}
\end{eqnarray}
The second term in the right side of the equation for $\dot{\phi}$
describes the nonlinear frequency shift of the standing wave. One
can see that, as the wave amplitude grows, its frequency goes down
\cite{only_standing}. This fact is important in the autoresonance
excitation scheme introduced below. Rescaling time, amplitude and
detuning,
\begin{equation}\label{def}
\tau=\varepsilon\omega_{1}t\;\,,\;\;\;B=\frac{k_{1}}{2\sqrt{\varepsilon}}A_{1}\;\,,\;\;\;\Delta=\frac{\delta}{\varepsilon\omega_{1}}\,,
\end{equation}
we rewrite Eqs. (\ref{averaged1}) in a scaled form:
\begin{eqnarray}
\dot{B}&=&B\sin(2\phi)\,,\nonumber\\
\dot{\phi}&=&\cos(2\phi)-B^{2}+\Delta\,, \label{main}
\end{eqnarray}
where the time derivatives are taken with respect to the slow time $\tau$. When $\Delta \lesssim{\cal O} (1)$,
the typical value of $B$ (for example the stable fixed point, see below) is ${\cal O} (1)$. Going back to Eq.
(\ref{def}), we see that, in the dimensional units, the parameter $k_{1}A_{1}\sim \varepsilon^{1/2} \ll 1$. As
in the leading order $\kappa\simeq\,k_{1}A_{1}$, this validates our assumption that $\kappa\ll\,1$.

Equations (\ref{main}) describe weakly-nonlinear
constant-frequency Faraday waves in the leading order in
$\varepsilon$. In the context of Faraday waves, Eqns. (\ref{main})
were first obtained by Miles \cite{Miles1,Miles2}, though he
derived them in a different way, working with the Lagrangian of
the fluid. In the sub-leading order in $\varepsilon$, additional
nonlinear terms appear \cite{Miles3,douady,milner,decent}, which
will not be considered here.

Equations (\ref{main}) can be rewritten in a Hamiltonian
form if we introduce the action and angle variables
$I=B^{2}/2\;\;\mbox{and}\;\;\phi$:
\begin{eqnarray}
\dot{I}&=&-\frac{\partial\,H}{\partial\phi}=2I\sin(2\phi)\,,\nonumber\\
\dot{\phi}&=&\frac{\partial\,H}{\partial\,I}=\cos(2\phi)-2I+\Delta\,,
\label{main1}
\end{eqnarray}
where the Hamilton's function is
\begin{equation}\label{hamiltonian}
H(I,\phi)=I\cos(2\phi)-I^{2}+\Delta\,I\,,
\end{equation}

\begin{figure}[ht]
\includegraphics[width=7cm,clip=]{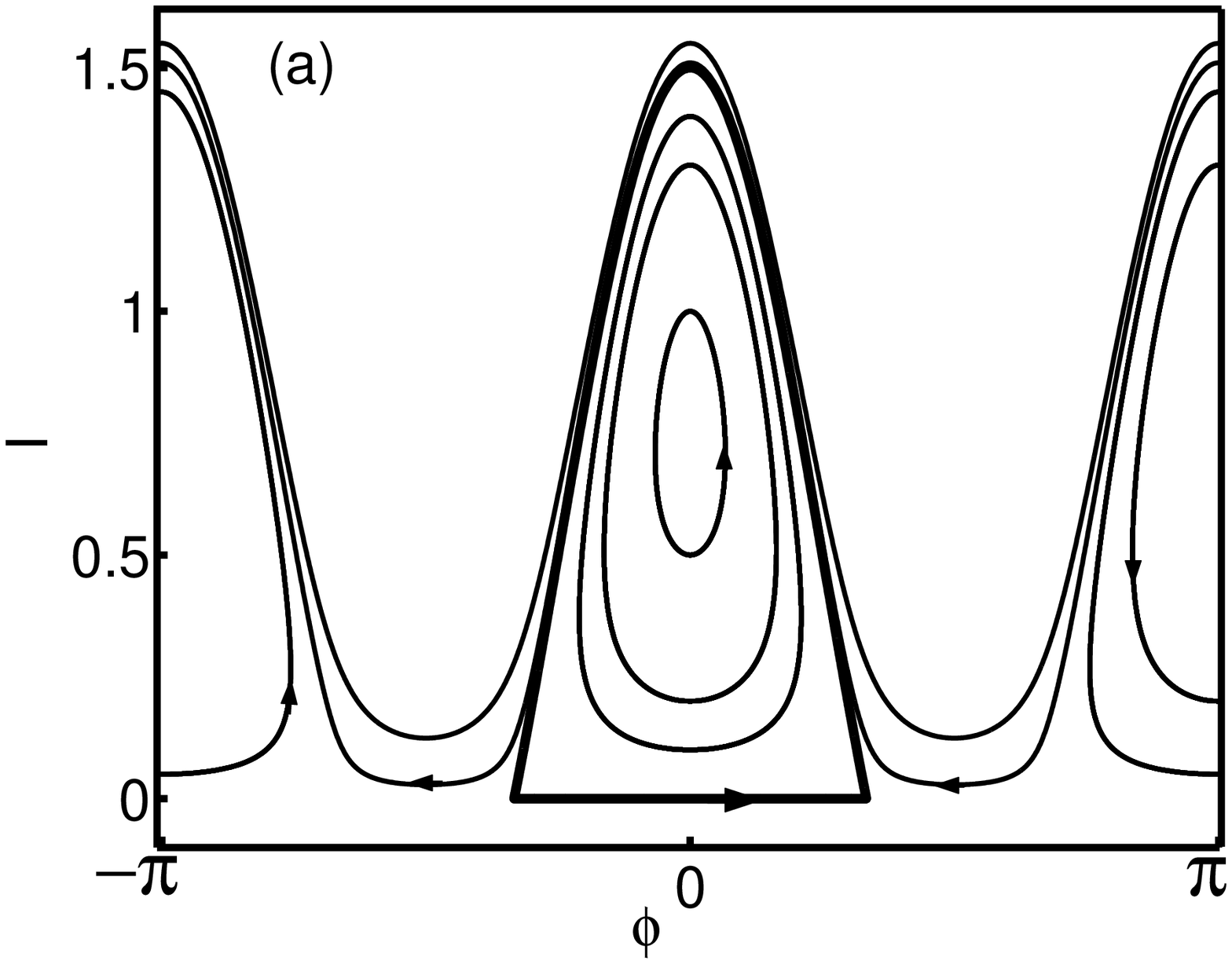}
\includegraphics[width=7cm,clip=]{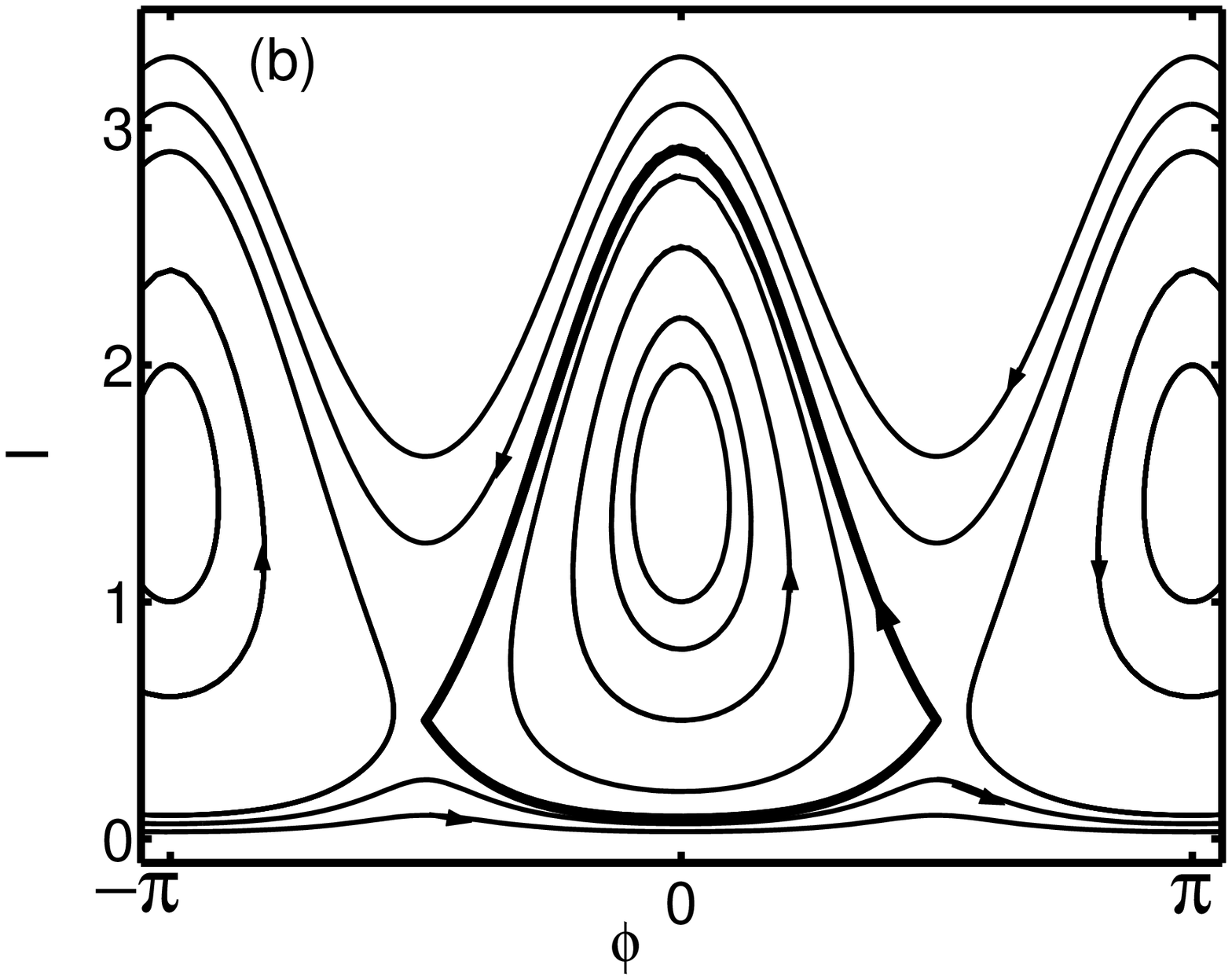}
\caption{The phase portrait of the inviscid constant-frequency system (\ref{main1}) with detuning $\Delta=0.5$
(a) and $\Delta=2$ (b). Phase locking occurs inside the regions limited by the separatrix (denoted by the thick
line). The saddle points are $[I^*,\phi^*]=[0,\pm\arccos(-\Delta/2)]$ in case (a) and $[1/2,\pm\pi/2]$ in case
(b). The phase portrait is periodic in $\phi$ with period $\pi$.} \label{pdiagnonzero}
\end{figure}

The fixed points of this dynamical system are determined by the
value of the scaled detuning $\Delta$ \cite{Bogoliubov,Struble}:
\begin{itemize}
\item[(a)] $\Delta<-1$. No fixed points.
\item[(b)] $-1<\Delta<1$. Three fixed points: an elliptic point
$[I_{*},\phi_{*}]=[(1+\Delta)/2,0]$ and two saddle points
$[I_{*},\phi_{*}]=[0, \pm\arccos(-\Delta)/2]$.
\item[(c)]
$\Delta>1$. Two fixed points: the same elliptic point
$[I_{*},\phi_{*}]=[(1+\Delta)/2,0]$ as in case (b), and a saddle
point $[I_{*},\phi_{*}]=[(\Delta-1)/2,\pi/2]$.
\end{itemize}
Figure \ref{pdiagnonzero} shows the phase plane ($\phi$,$I$) in the cases of $0<\Delta<1$ and $\Delta>1$.  The
phase portrait is periodic in $\phi$ with period $\pi$.  In the case of $0<\Delta<1$, the separatrix is formed
by the curve $I=\Delta+\cos(2\phi)$ and the straight line $I=0$. In the case of $\Delta>1$, the separatrix is
formed by the curves $I=[\cos(2\phi)+\Delta-\sqrt{\alpha}]/2$ and $I=[\cos(2\phi)+\Delta+\sqrt{\alpha}]/2$,
where $\alpha= \cos^{2}(2\phi)+2\Delta\cos(2\phi)+2\Delta\,-1$. Notice that the maximum possible amplitude of
phase-locked oscillations is achieved at a nonzero detuning from the exact linear resonance, like in many other
instances of nonlinear resonance.

As the Hamilton's function (\ref{hamiltonian}) is a constant of
motion, the system is integrable. In particular, one can find the
``nonlinear period": the period of motion along a closed
trajectory in the phase plane. Denoting the
constant Hamilton's function as $H_{0}$, 
we obtain
\begin{equation}\label{period}
T_{nl}=2\int_{\phi_{-}}^{\phi_{+}}\frac{d\phi}{\left\{\left[\Delta+\cos(2\phi)\right]^{2}-4H_{0}\right\}^{1/2}}\,,
\end{equation}
where $\phi_{\pm}=\pm\arccos(2\sqrt{H_{0}}-\Delta)/2$. For a zero
detuning, and initial conditions very close to the fixed point
$I_{*}=1/2$ and $\phi_{*}=0$ (so that $H_{0}\simeq\,1/4$), we
obtain $T\simeq\pi$. This corresponds to small harmonic
oscillations around the elliptic fixed point. In the physical
units the period of small oscillations is
$T_{nl}^{ph}\simeq\pi/(\varepsilon \omega_1)$, that is much longer
than the wave period.


\subsection{Low-viscosity fluid}
Taking into account a weak damping of the wave amounts to adding a
linear damping term $2\gamma\dot{\eta}_{1}$ to the left side of
Eq. (\ref{motion}), where $\gamma$ is defined in terms of the rate
of loss of mechanical energy due to dissipation,
\cite{Landau2}. The incorporation of only a \textit{linear}
damping term requires that $\gamma/\omega_{1}\ll\,1$, so that the
damping is treated perturbatively. The specific damping mechanisms
which contribute to the value of damping rate $\gamma$ are the
bulk viscosity \cite{Landau2}, dissipation in the vicinity of the
fixed walls \cite{Miles4}, dissipation at the free surface
(especially if contaminated) \cite{Miles4} and contact line
damping \cite{christiansen} (see, \textit{e.g.} Ref.
\cite{christiansen} for a review). In practice, one can interpret
the damping rate as a phenomenological term, and determine it from
a comparison with experiment.


\begin{figure} [ht]
\includegraphics[width=7cm,clip=]{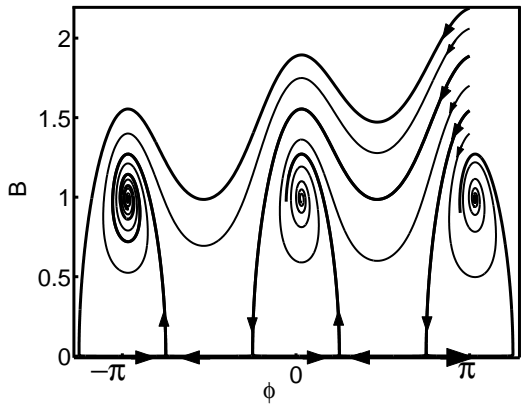}
\caption{The phase portrait of the constant-frequency system (\ref{mainvisc}) with a zero detuning and scaled
damping rate $\Gamma=0.2$. The thick lines mark the separatrices, which pass through the saddle points
$[B_*,\phi_*]=[0,\pm\pi/4]$. The phase portrait is periodic in $\phi$ with period $\pi$.} \label{pdiagvisc}
\end{figure}

\begin{figure}[ht]
\includegraphics[width=7cm,clip=]{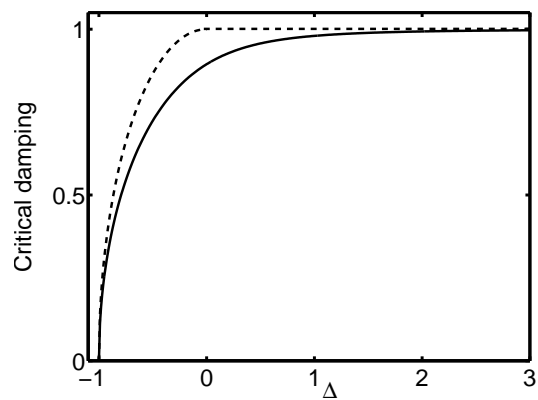}
\caption{The solid line shows the critical scaled damping rate $\Gamma_{cr}$ as a function of the scaled
detuning $\Delta$ [Eq. (\ref{gammacrit})]. For $\Gamma>\Gamma_{cr}$ (the over-damped case) a stable node is
obtained, while for $\Gamma<\Gamma_{cr}$ a stable focus is obtained. The dashed line shows the maximum value of
$\Gamma$ for which a nontrivial stable fixed point still exists. $\Gamma_{cr}$ is always below this maximum
value.} \label{critdamp}
\end{figure}

\begin{figure}[ht]
\includegraphics[width=7cm,clip=]{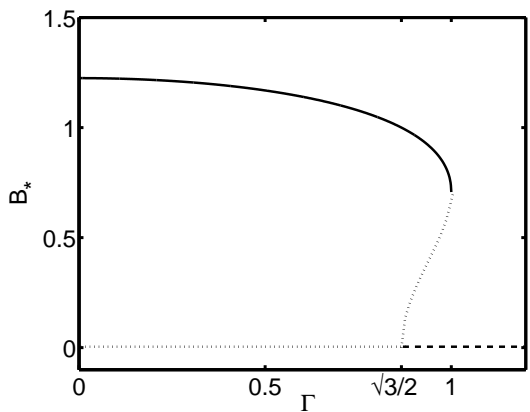}
\caption{A combined bifurcation diagram of the constant-frequency system (\ref{mainvisc}). Shown is the
fixed-point amplitude $B_{*}$ versus $\Gamma$ for $\Delta=0.5$. The solid line shows the stable focus or stable
node (depending on $\Gamma$), the dashed line shows the stable node \textit{and} the saddle point [as described
in case (c)], and the dotted lines show the unstable fixed points (saddle points and unstable focus). The
bifurcation occurs at $\Gamma=\sqrt{1-\Delta^{2}}$, where the stable node
$[B_*,\phi_*]=[0,(1/2)\arccos(-\Delta)]$ becomes a saddle point. In the region $\sqrt{1-\Delta^{2}}<\Gamma<1$
both the stable focus, and the stable node exist.} \label{bifurdel05}
\end{figure}

Including the linear damping term in the first of Eqs.
(\ref{main}), we obtain:
\begin{eqnarray}
\dot{B}&=&B\sin(2\phi)-\Gamma\,B\,,\nonumber\\
\dot{\phi}&=&\cos(2\phi)-B^{2}+\Delta\,, \label{mainvisc}
\end{eqnarray}
where $\Gamma=\gamma/(\varepsilon\omega_{1})>0$ is a dimensionless damping rate. It follows from the first of
Eqs. (\ref{mainvisc}), that non-trivial fixed points $B_* \neq 0$ can exist only when $\Gamma<1$, that is for a
small enough viscosity. In this case, any trajectory on the phase plane of the system (except for trajectories
with a zero measure) converges to a stable focus, see Fig. \ref{pdiagvisc}. This is in contrast to the inviscid
case, where starting from initial conditions outside the separatrix leaves the trajectory phase-unlocked. The
fixed points $[B_{*},\phi_{*}]$ of Eqs. (\ref{mainvisc}) are determined by the values of $\Delta$ and $\Gamma$.
Since Eqs. (\ref{mainvisc}) [and Eqs. (\ref{main})] are invariant under the transformation $B \to -B$, one needs
to consider only fixed points with positive amplitudes. Let $\sigma=\sqrt{1-\Gamma^{2}}$ and
$\xi=\sqrt{1-\Delta^{2}}$. Let us also denote the critical damping rate
\begin{equation}\label{gammacrit}
\Gamma_{cr}=\left[\frac{4}{5}-\frac{8\Delta}{25}\left(\Delta-\sqrt{\Delta^{2}+\frac{5}{4}}\right)\right]^{1/2}\,,
\end{equation}
which will appear shortly. There are four different cases:
\begin{itemize}
\item[(a)] $\Delta<-1$. No fixed points. \item[(b)] $-1\leq\Delta<0$. For $\Gamma>\xi$, $[0,\arccos(-\Delta)/2]$
is a stable node, and $[0,-\arccos(-\Delta)/2]$ is a saddle point. For $0\,\leq\,\Gamma\,\leq\,\xi$,
$[0,\pm\arccos(-\Delta)/2]$ are two saddle points, and $[(\sigma+\Delta)^{1/2}, \arcsin(\Gamma)/2]$ is a stable
fixed point. For $0\leq\Gamma<\Gamma_{cr}$ it is a stable focus, while for $\Gamma_{cr}<\Gamma\leq\xi$ it is a
stable node.

\item[(c)] $0\leq\Delta<1$. For $\Gamma\geq\,1$, $[0,\arccos(-\Delta)/2]$ is a stable node, and
$[0,-\arccos(-\Delta)/2]$ is a saddle point. For $\xi\,<\,\Gamma\,<\,1$, $[0,\arccos(-\Delta)/2]$ is a stable
node, $[0,-\arccos(-\Delta)/2]$ and $[(\Delta-\sigma)^{1/2},\pi/2-\arcsin(\Gamma)/2]$ are two saddle points, and
$[(\sigma+\Delta)^{1/2},\arcsin(\Gamma)/2]$ is a stable fixed point. For $\xi<\Gamma<\Gamma_{cr}$ it is a stable
focus, while for $\Gamma_{cr}<\Gamma<1$ it is a stable node. For $0\,\leq\,\Gamma\,\leq\,\xi$,
$[0,\pm\,\arccos(-\Delta)/2]$ are two saddle points, and $[(\sigma+\Delta)^{1/2},\arcsin(\Gamma)/2]$ is a stable
focus. \item[(d)] $\Delta \ge 1$. For $0\leq\Gamma\leq\,1$, $[(\sigma+\Delta)^{1/2},\arcsin(\Gamma)/2]$ is a
stable fixed point. For $0\leq\Gamma<\Gamma_{cr}$ it is a stable focus, while for $\Gamma_{cr}<\Gamma\leq\,1$ it
is a stable node, and $[(\Delta-\sigma)^{1/2},\pi/2-\arcsin(\Gamma)/2]$ is a saddle point.
\end{itemize}
Figure \ref{critdamp} shows two characteristic values of the
scaled damping $\Gamma$ as functions of $\Delta$. The first one is
$\Gamma_{cr}$ from Eq. (\ref{gammacrit}). The second one is the
maximum value of $\Gamma$ for which a nontrivial stable fixed
point still exists. For $-1<\Delta<0$ this maximum value is equal
to $\sqrt{1-\Delta^{2}}$, while for $\Delta\geq\,0$ it is equal to
$1$. To conclude the brief review of the constant-frequency
theory, we notice that the dependence of $B_*$ on $\Gamma$
exhibits a pitchfork bifurcation. Figure \ref{bifurdel05} shows
the bifurcation diagram in case (c) for $\Delta=0.5$.

\section{Chirped Faraday waves in an inviscid fluid}
\subsection{Governing equations, phase portrait, criteria and numerical examples}
Now let the vibration frequency be time-dependent (chirped):
$\omega=\omega(t)$.  In general, the dimensionless parameter
$\varepsilon=\omega^2 a_0/g$ will also become time-dependent. For
simplicity, we shall assume that $a_0$ also varies in time so that
$\varepsilon=const$ \cite{constamp}. Our objective is to keep a
Faraday wave close to resonance in spite of its nonlinear
frequency shift, so as to achieve a persistent growth of the wave
amplitude. Like in other autoresonance schemes, the exact form of
the function $\omega(t)$ is unimportant if this function satisfies
\textit{three} criteria:

\begin{enumerate}
\item{The chirp sign coincides with the sign of the nonlinear
frequency shift of the wave. For the standing Faraday waves
$\omega(t)$ should \textit{decrease} for the wave amplitude to
increase.}

\item{The frequency chirp rate must be sufficiently small, so that
the phase portrait of the system evolves adiabatically:
$|\dot{\omega(t)}|\,T_{nl}\ll\omega(t)$, where $T_{nl}$ is the
characteristic nonlinear period, see Eq. (\ref{period}).}

\item{The dynamic frequency mismatch, which we define as the
absolute value of the increment of the vibration frequency during
one nonlinear period, $|\omega(t+T_{nl})-\omega(t)|$, should be
small compared with the inverse nonlinear period, $T_{nl}^{-1}$.
In physical units, this yields
$\mu/(\varepsilon\omega_{1})^{2}\ll\,1$.}
\end{enumerate}

Criteria 1 and 2 have appeared in previous works on autoresonance
\cite{Friedland1}, while criterion 3 is new. We shall see shortly
that, in the problem of parametric autoresonance,  criterion 3 is
more restrictive than criterion 2.

The derivation of the equation of motion for the primary mode
amplitude, for a slowly time-dependent driving frequency, goes
along the same lines as in the case of a constant driving
frequency. The resulting equation is (compare with Eq.
(\ref{motion})]:
\begin{equation}\label{motionchirp}
\ddot{\eta}_{1}+\frac{1}{2}k_{1}^{2}(5\dot{\eta}_{1}^{2}\eta_{1}-3\omega_{1}^{2}\eta_{1}^{3})+\omega_{1}^{2}\left[1+4\varepsilon\cos(2\Phi(t))\right]\eta_{1}=0\,,
\end{equation}
where $\Phi (t)=\int_0^t\omega(t^{\prime}) \, dt^{\prime}$. If
$\varepsilon$ is small, and the chirp rate is slow on the time
scale of the wave period, one can again use, close to the
parametric resonance, the method of averaging \cite{Khain}. For
concreteness, we assume in this work a constant chirp rate $\mu$:
\begin{equation}\label{omegatime}
\omega(t)=\omega_{1}-\mu\,t\,,
\end{equation}
so that $\Phi (t)=\omega_{1} t - \mu t^2/2$. Introducing a scaled
chirp rate $m=\mu/(\omega_{1}\varepsilon)^{2}$, and the same
scaled time $\tau$ and amplitude $B$ as before [see Eq.
(\ref{def})], we arrive at the following scaled equations:
\begin{eqnarray}
\dot{B}&=&B\sin(2\phi)\,,\nonumber\\
\dot{\phi}&=&\cos(2\phi)-B^{2}+m\tau, \label{chirp}
\end{eqnarray}
where now $\phi (t) = \mu t^2/2+\phi_1(t)$, and the
differentiation is done with respect to $\tau$. In the
action-angle variables we obtain:
\begin{eqnarray}
\dot{I}&=&2I\sin(2\phi)\,,\nonumber\\
\dot{\phi}&=&\cos(2\phi)-2I+m\tau\,. \label{chirp1}
\end{eqnarray}
Once $B(t)$ and $\phi(t)$ are found, one can immediately
reconstruct the standing way profile $\eta(x,t)$ by using the
Ansatz
$\eta_{1}(t)=A_{1}(t)\cos\left[\omega_{1}t+\phi_{1}(t)\right]$ and
the ``enslaving relation" (\ref{enslaving}) in the second of
equations (\ref{waveprof}). Therefore, in the rest of the paper we
shall focus on Eqs. (\ref{chirp1}) which coincide, up to notation,
with those obtained by Khain and Meerson \cite{Khain}, who
investigated parametric autoresonance in a nonlinear oscillator.
Equations similar to (\ref{chirp1}) also appear in the problem of
the second-harmonic autoresonance in an \textit{externally} driven
oscillator \cite{Lazar7}.
\begin{figure}[ht]
\includegraphics[width=7cm,clip=]{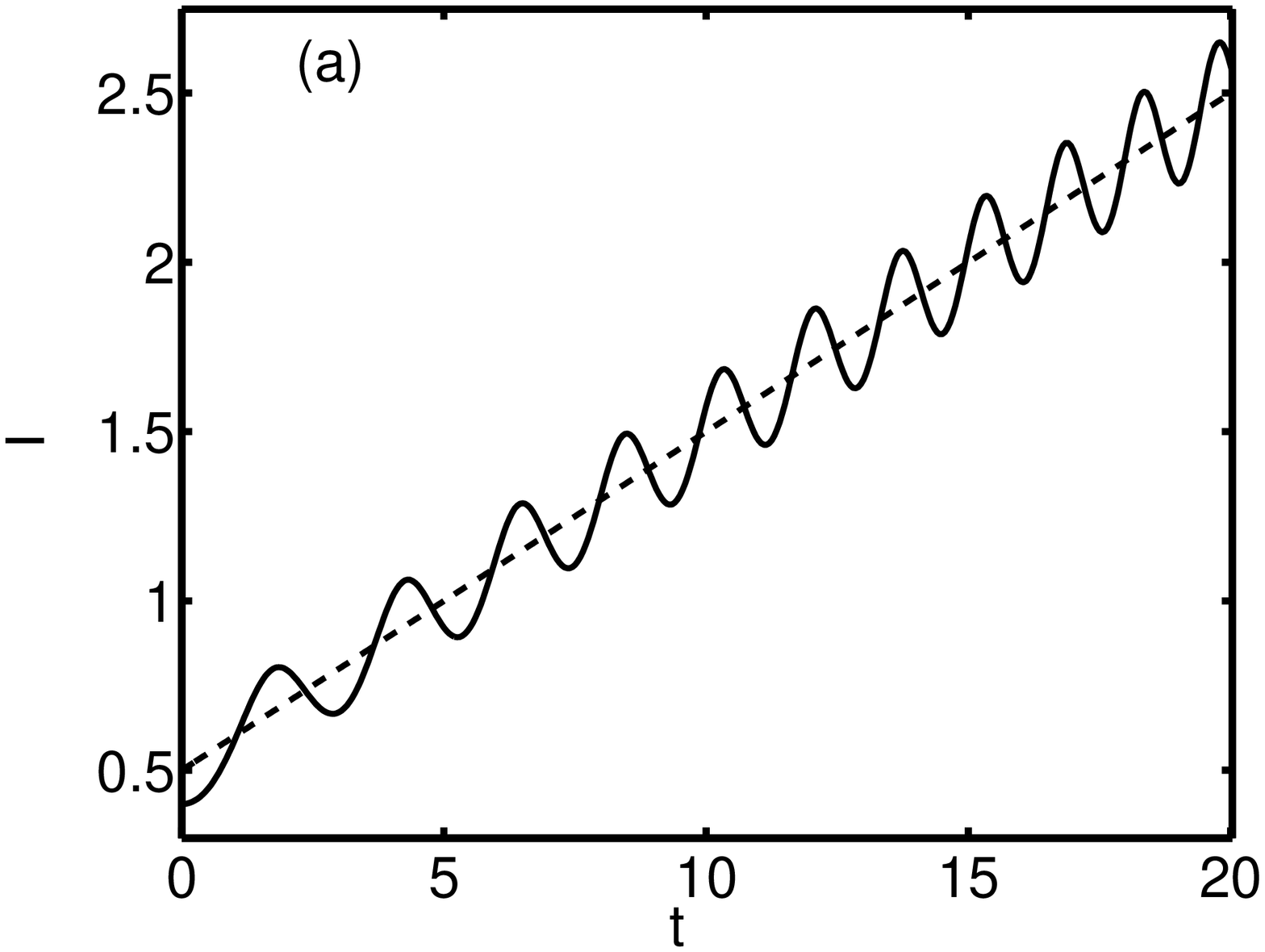}
\includegraphics[width=7cm,clip=]{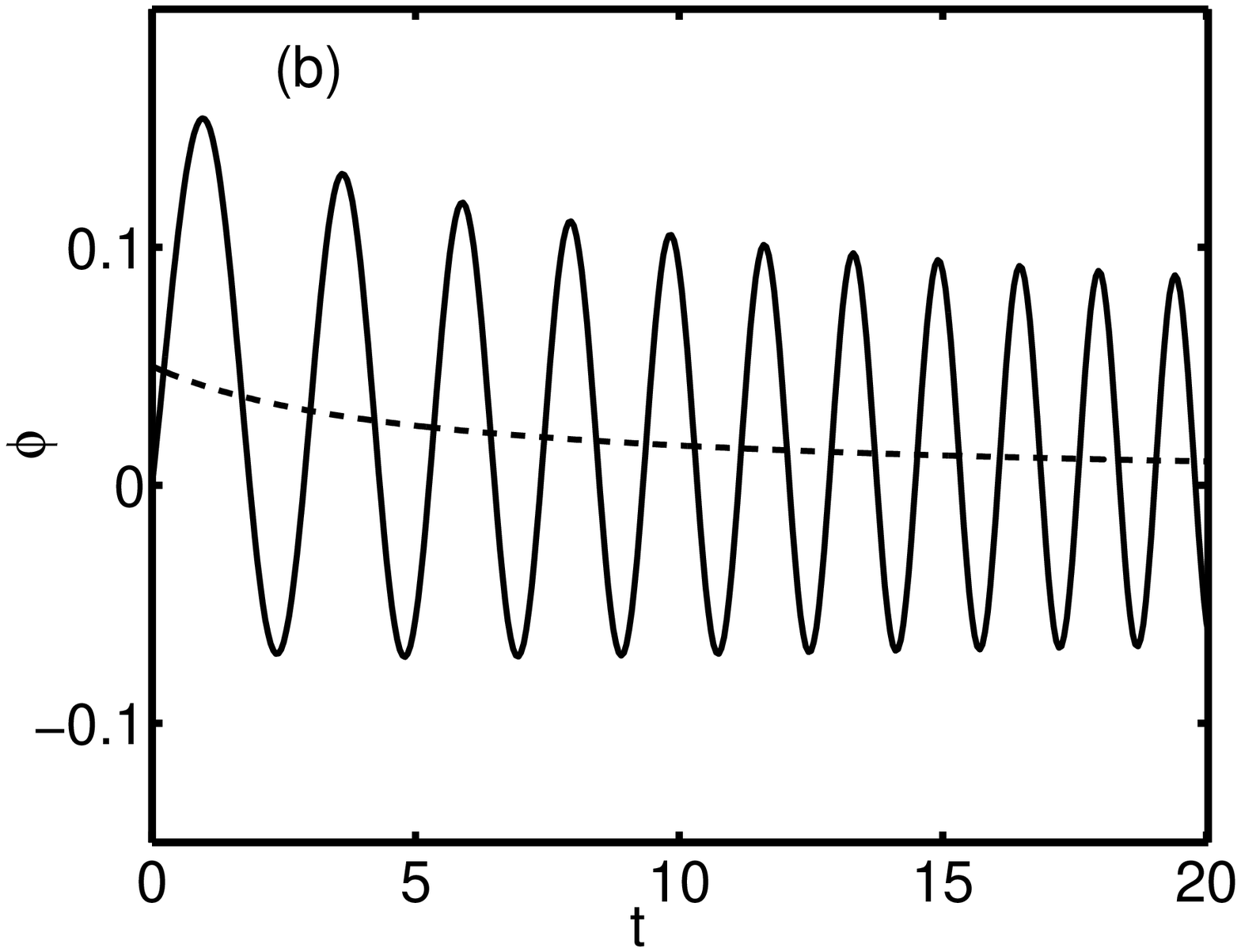}
\caption{Parametric autoresonance in an inviscid fluid. Shown are the action $I$ (a) and the phase $\phi$ (b) as
functions of time. The parameters are $m=0.2$, $I(0)=0.4$, $\phi(0)=0$, and a zero initial detuning. The solid
lines shows numerical solutions. The dashed lines show the trends $I_{*}(t)$, given by Eq. (\ref{istar}) (a) and
$\phi_{*}(t)$, given by Eq. (\ref{phistar}) (b).} \label{chirptrend}
\end{figure}

The Hamilton's function of the system (\ref{chirp1}) is time
dependent:
\begin{equation}\label{hamchirp}
H(I,\phi)=I\cos(2\phi)-I^{2}+ m\tau I\,,
\end{equation}
so $H$ is not a constant of motion anymore. In the following we
shall use $t$ for the slow time $\tau$.

\begin{figure}[b]
\includegraphics[width=7cm,clip=]{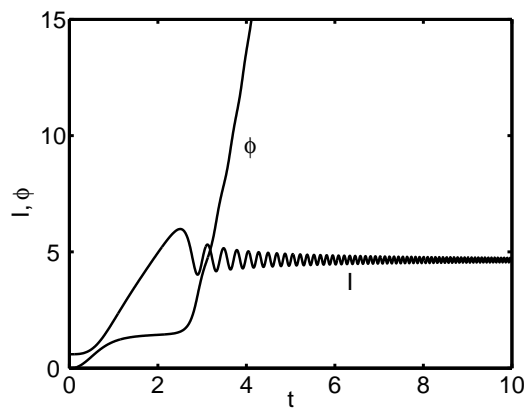}
\caption{
Breakdown of parametric autoresonance in an inviscid fluid in the case of $m=5.5>m_{cr}$. 
The parameters are $m=5.5$, $I(0)=0.6$, $\phi(0)=0$, and a zero initial detuning. Shown are the action $I(t)$
and the phase $\phi(t)$ found numerically.} \label{arandnar}
\end{figure}

Figure \ref{chirptrend} shows an example of parametric
autoresonance: a persistent phase locking and a systematic growth
of $I$ with time, with some oscillations on top of the systematic
growth.  Here the scaled chirp rate $m$ is less than some critical
value $m_{cr}$ for these initial conditions. Figure \ref{arandnar}
illustrates breakdown of autoresonance, observed when $m>m_{cr}$.
As in other instances of autoresonance, a theory of parametric
autoresonance appeals to the constant-frequency theory. Comparing
Eqs. (\ref{chirp1}) and (\ref{main1}), one can see that the term
$m t$ plays the role of an effective (time-dependent) detuning.
Therefore, when the chirp rate is small, $m\ll\,1$, the phase
portrait of the system almost coincides with that of the
autonomous equations (\ref{main1}) (see Fig. 1), except that now
it changes with time according to the current value of the
effective detuning. The change of the phase portrait is
adiabatically slow, except at times $t\simeq -1/m$ and $1/m$
(corresponding to $\Delta=-1$ and $1$, respectively), when
bifurcations occur. One consequence of the adiabatic evolution is
that Eqs. (\ref{chirp1}) have ``quasi-fixed" points. The most
important \textit{stable} quasi-fixed point $[I_*(t),\phi_*(t)]$
can be found by assuming that $\phi_*(t) \ll 1$, and that it
varies with time slowly. Then the second of Eqs. (\ref{chirp1})
yields, in the leading order,
\begin{equation}\label{istar}
I_{*}\simeq\frac{1}{2}\left(1+mt\right)\,.
\end{equation}
Substituting this into the first of Eqs. (\ref{chirp1}), we obtain
\begin{equation}\label{phistar}
\phi_{*}\simeq\frac{m}{4(1+mt)}\,.
\end{equation}
The stable quasi-fixed point, or \textit{trends} (\ref{istar}) and
(\ref{phistar}), previously found by Khain and Meerson
\cite{Khain} (see also Ref. \cite{Lazar7}), are the essence of
parametric autoresonance. Shown in Fig. \ref{chirptrend} are
$I(t)$ and $\phi(t)$ found numerically, and the trends
(\ref{istar}) and (\ref{phistar}). The trend (\ref{istar})
corresponds to a steady growth of the wave amplitude: $B_*(t)=[2
I_*(t)]^{1/2}\simeq (mt+1)^{1/2}$. The important phase trend
(\ref{phistar}) was overlooked in Ref. \cite{Lazar7}. Notice that,
at scaled time $t \gg 1$, the phase trend $\phi_{*}\simeq 1/(4 t)$
becomes independent of the chirp rate $m$. Importantly, for the
expressions (\ref{istar}) and (\ref{phistar}) to be valid, one can
either demand $m \ll 1$, or go to long times: $t \gg 1$.
Therefore, the stable quasi-fixed point keeps its meaning, at long
times, even at finite (non-small) $m$.

We found that, surprisingly, \textit{unstable} quasi-fixed points
of the chirped system also play an important role in the dynamics.
The unstable points are analogs of the constant-frequency saddle
points discussed in the previous section [see the text following
Eq. (\ref{hamiltonian})]. To find the locations of the unstable
quasi-fixed points in the leading order, one can simply replace
the detuning $\Delta$ by $m t$. Therefore, on the time interval
$0<t<1/m$, there are two saddle quasi-fixed points
$[I_{*},\phi_{*}]\simeq\,[0, \pm(1/2)\arccos(-mt)]$. These points
disappear at $t\simeq 1/m$, and a new saddle points appears:
$[I_{*},\phi_{*}]\simeq\,[(m t-1)/2, \pi/2)]$. These expressions
(including the boundaries of the corresponding time intervals) are
valid in the leading order in $m \ll 1$. Higher-order corrections
can be also calculated.

Now we are in a position to discuss criteria 2 and 3 for
parametric autoresonance in this system. For a constant chirp rate
$\mu$ [see Eq. (\ref{omegatime})], criterion 2 can be written, in
the physical units, as $\mu (\varepsilon \omega)^{-1} < \omega$,
or $\mu < \varepsilon \omega^2$. Now, the dynamic frequency
mismatch, acquired by the chirped system during time $T_{nl}$, can
be estimated as $\mu T_{nl} \sim \mu/(\varepsilon \omega)$.
Criterion 3 demands that this quantity be small compared to
$T_{nl}^{-1} \sim \varepsilon \omega$, which yields $\mu <
\varepsilon^2 \omega^2$. As $\varepsilon$ is small, criterion 3 is
more restrictive than criterion 2. In the scaled variables,
criterion 3 has the form of $m < 1$, as can be expected from the
form of scaled Eqs. (\ref{chirp1}).  The inequalities here are
written up to numerical factors which depend on the initial
conditions, see below.

\begin{figure}[ht]
\includegraphics[width=7cm,clip=]{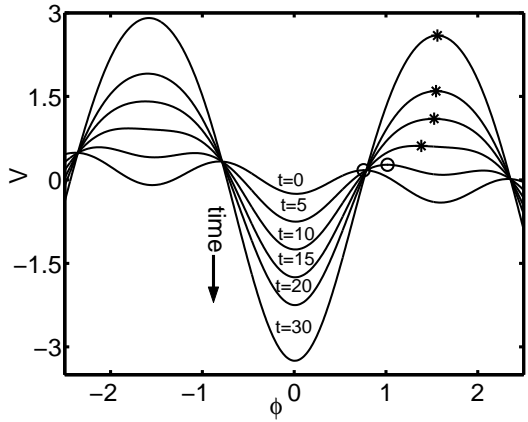}
\caption{The time-dependent potential $V(\phi,t)$, Eq. (\ref{pot}), is shown as a function of $\phi$ for several
consecutive times starting from $t=0$. The scaled chirped rate $m=0.1$. The potential well at
$\phi\simeq\phi_{*}(t)$ deepens with time starting from $t=0$. At $0<t<1/m$ there is a potential barrier at
$\phi\simeq(1/2)\arccos(-mt)$ (denoted by the circle), which disappears at $t\simeq1/m$. At $t\simeq1/m$ a new
potential barrier appears close to $\phi=\pi/2$ and heightens with time (denoted by the asterisk).}
\label{potfig}
\end{figure}

One more convenient description of the chirped system can be
achieved if we rewrite Eqs. (\ref{chirp1}) as a second order
equation for the phase:
\begin{equation}\label{diffphi}
\ddot{\phi}+2mt\sin(2\phi)+\sin(4\phi)-m=0\,,
\end{equation}
or
\begin{equation}
\ddot{\phi}+\frac{\partial\,V(\phi,t)}{\partial\phi}=0\,,
\end{equation}
where we have introduced a time-dependent potential
\begin{equation}\label{pot}
V(\phi,t)=-\frac{1}{4}\cos(4\phi)-mt\cos(2\phi)-m\phi\,.
\end{equation}
This suggests new canonical variables $\phi$ and $u=\cos (2 \phi)-2 I + m t$, so that in the new time-dependent
Hamiltonian, $H(\phi,u,t)=u^{2}/2+V(\phi,t)$, there is a clear separation between the potential energy and the
kinetic energy. The new Hamiltonian describes a ``particle" of a unit mass and velocity $u=\dot{\phi}$, moving
in a time-dependent potential $V$. This picture is useful for a qualitative analysis of the dynamics of the
``particle" when $m$ is small, so the potential slowly varies in time, see Fig. \ref{potfig}. In the variables
$u, \phi $ the stable quasi-fixed point becomes (approximately) $[0,(m/4) (1+m t)^{-1}]$, while the saddle
points are $[0,\pm (1/2)\arccos(-mt)]$ at $0<t<1/m$, and $[0,\simeq \pm\,\pi/2]$ at $t>1/m$. We shall see
shortly that each of the unstable points $[0, (1/2)\arccos(-mt)]$ and $[0,\simeq \pi/2]$ plays an important role
in this system.

Let us consider two typical cases of parametric autoresonant
excitation of a Faraday wave. In the first case one first excites
the wave at a constant frequency, so that the initial values of
the action and phase are in the vicinity of the stable fixed
point. Then, upon slowly reducing the driving frequency, one keeps
the phase locked, as our ``particle" oscillates in a potential
well which slowly deepens with time, see Fig. \ref{potfig}. In the
second case one starts the autoresonant driving from an almost
zero wave amplitude. Here the saddle point
$[I_{*},\phi_{*}]\simeq\,[0,(1/2)\arccos(-mt)]$ plays an important
role. Indeed, the stable manifold of this quasi-fixed point is
along the $\phi$ axis. Therefore, no matter what the initial phase
is, the phase approaches, on a time scale ${\cal O}(1)$, the
saddle point. The unstable manifold of this saddle point is along
the $I$ axis, so $I(t)$ will grow with time. Still, if $I(t=0)$ is
small enough, $I(t)$ remains small during this time scale ${\cal
O}(1)$. Therefore, phase locking is always achieved at this stage,
so the time interval $0<t<1/m$ can be called the ``trapping
stage". Later on $I(t)$ grows significantly but, as we found
numerically, the ``particles" remain inside the (slowly expanding)
separatrix $I=\cos(2\phi)+mt$. As a result, the phase starts to
perform large-amplitude oscillation around the stable quasi-fixed
point, and phase locking persists.


One more alternative description of the system of equations
(\ref{chirp}) is in terms of the complex amplitude
$\psi(t)=B(t)\exp[i\phi(t)]$:
\begin{equation}\label{NLS}
i\psi_{t}+\psi^{*}-(|\psi|^{2}-mt)\psi=0\,,
\end{equation}
where the subscript $t$ denotes differentiation with respect to
the slow time. 
The long-time behavior of $I_*$ can be obtained by looking at the
asymptotic solutions of Eq. (\ref{NLS}) at $t\rightarrow\infty$.
For a solution such that $|\psi|$ grows with time like a power
law, the leading terms are those in the parentheses. This
immediately yields
\begin{equation}\label{leadingB}
|\psi(t)|\simeq (mt)^{1/2}\,,
\end{equation}
which corresponds to the leading term (when $mt\gg\,1$) of Eq. (\ref{istar}), and describes a phase-locked wave
[here $\phi$ stays close to zero, see Eq. (\ref{phistar})]. On the contrary, if $|\psi(t)|$ remains bounded and
small, the first term of Eq. (\ref{NLS}) is balanced by the last one, and we obtain
\begin{equation}\label{limitphi}
\psi(t)=\psi_{0}\exp\left(i mt^{2}/2\right)\,,
\end{equation}
where $\psi_{0} \equiv |\psi_{0}| \exp(i \phi_{0})=const.$ This
solution corresponds to an unlocked phase $\phi(t) =
\phi_0+mt^2/2$ and a constant amplitude $|\psi_{0}|$. Of course,
the phase of the wave $\phi_1(t)=\phi(t)-mt^{2}/2$, which is
defined by the Ansatz
$\eta_{1}(t)=A_{1}(t)\cos\left[\omega_{1}t+\phi_{1}(t)\right]$
(where $t$ is the physical time), stays constant in this regime,
and is equal to $\phi_{0}$.

\begin{figure}[ht]
\includegraphics[width=7cm,clip=]{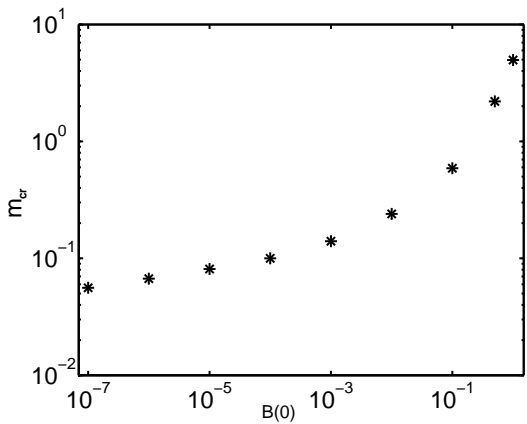}
\caption{The critical chirp rate $m_{cr}$ as a function of the
initial amplitude $B(0)$ for $\phi(0)=0$.} \label{mcritamp}
\end{figure}

We determined numerically, for several typical classes of initial
conditions, the critical value of $m$, $m=m_{cr}$, which separates
the phase locking regime from the phase unlocking regime
\cite{alternative}. At a fixed initial phase $\phi(t=0)=0$,
$m_{cr}$ grows with the initial amplitude $B(t=0)$, at least until
the scaled amplitude becomes of order unity, see Fig.
\ref{mcritamp}. This result agrees with those obtained by Fajans
\textit{et. al.} (see Fig. 3 in Ref. \cite{Lazar7}) who presented
them in terms of $\varepsilon_{cr}$ versus $\mu$. When starting
from the stable quasi-fixed point: $I(t=0)=1/2$ and $
\phi(t=0)=0$, we found that $m_{cr}\simeq\,4.963$.

\begin{figure}[ht]
\includegraphics[width=7cm,clip=]{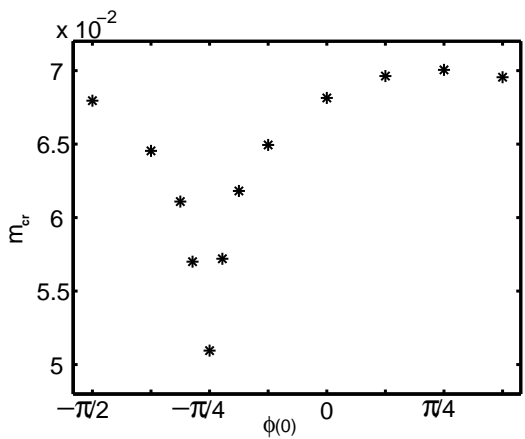}
\caption{The critical chirp rate $m_{cr}$ as a function of the
initial phase $\phi(0)$, for a very small initial amplitude
$B(0)=10^{-6}$.} \label{mcritical}
\end{figure}

We also found $m_{cr}$ as a function of the initial phase, for a
given, and very small, initial amplitude, see Fig.
\ref{mcritical}. This dependence is relatively weak. The largest
$m_{cr}$ is obtained for $\phi=\pi/4$, the smallest one for
$\phi=-\pi/4$.

\begin{figure}[ht]
\includegraphics[width=7cm,clip=]{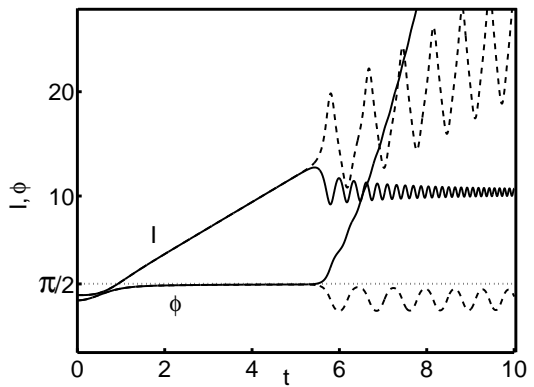}
\caption{Autoresonance and its breakdown for $m$ very close to
$m_{cr}\simeq 4.963$. The initial conditions are $\phi(t=0)=0$ and
$I(t=0)=1/2$. The dashed lines show the phase-locked solutions for
the phase and action for $m$ just below $m_{cr}$, while the solid
lines show the phase-unlocked solutions for $m$ just above
$m_{cr}$.} \label{maxmu1}
\end{figure}

What is the signature of the special case $m=m_{cr}$? At
$m<m_{cr}$ the phase oscillates in time. As $m$ approaches
$m_{cr}$ from below, the onset of the phase oscillations is
delayed more and more, see Fig. \ref{maxmu1}. Now, at $m>m_{cr}$
the phase initially changes slowly and then rapidly escapes to
infinity. As $m$ approaches $m_{cr}$ from above, the point of
rapid departure of the phase is delayed more and more, as shown in
Fig. \ref{maxmu1}. This suggests that, in the special case
$m=m_{cr}$ the phase neither oscillates around a trend, nor
departs from it. Instead, the phase monotonically approaches
$\pi/2$. Looking at the effective potential, shown in Fig.
\ref{potfig}, one realizes that, in the special case $m=m_{cr}$,
our ``particle" neither oscillates in the potential well (which
would correspond to phase locking), nor escapes from the well
(which would correspond to phase unlocking). Instead, the
``particle" lands, at $t=\infty$, on the peak of the potential at
$\phi=\pi/2$. In the following we shall find the asymptotic form
of this special trajectory.

\subsection{Perturbative solutions}
In this subsection we present three perturbative analytic
solutions which illustrate the basic features of parametric
autoresonance: persistent resonant growth, capture into resonance
and the limiting trajectory which separates between phase locking
and unlocking. In each of the three cases, a local analysis around
one of the quasi-fixed points of the system is required.

\subsubsection{In the vicinity of the stable quasi-fixed point}

\begin{figure}[b]
\includegraphics[width=7cm,clip=]{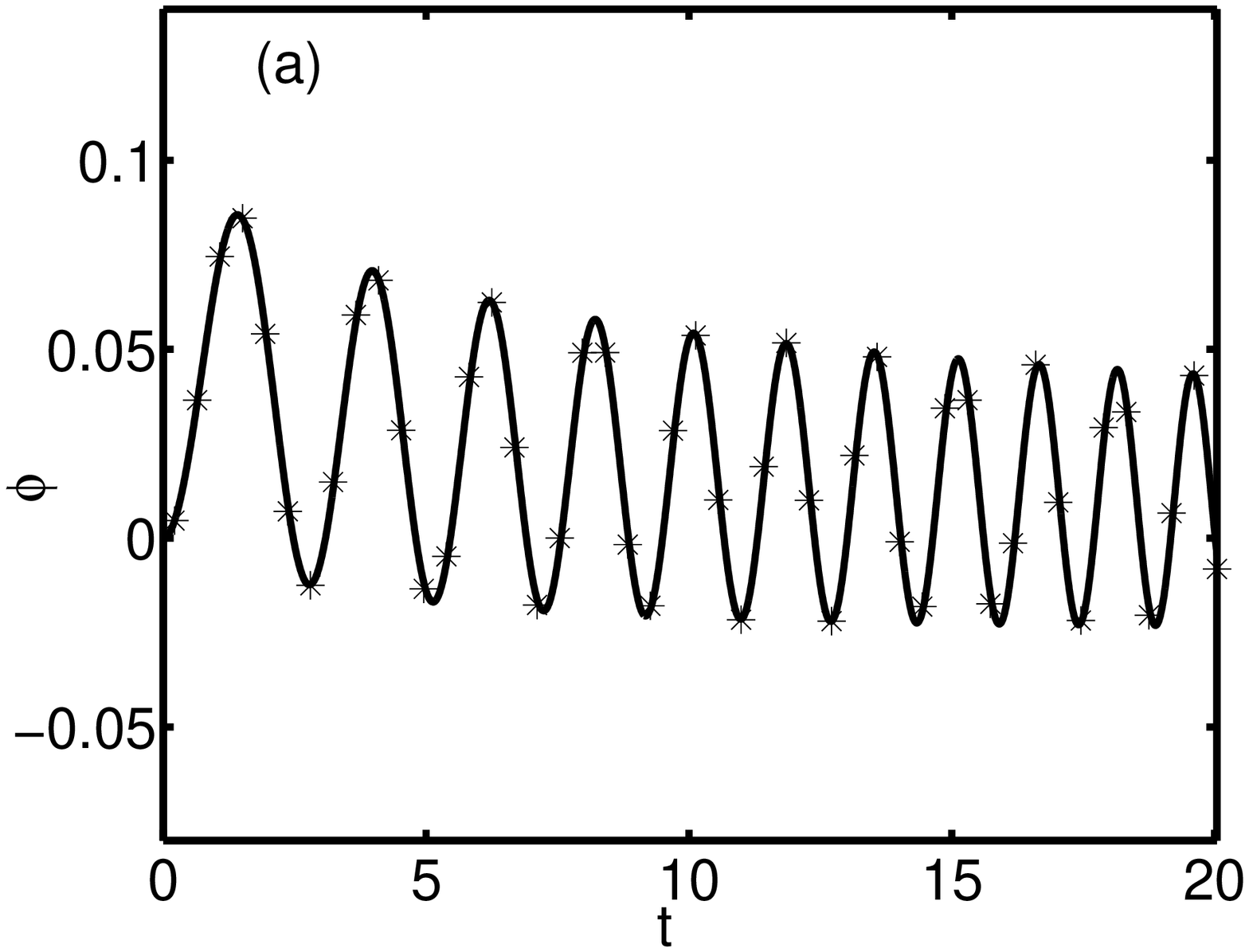}
\includegraphics[width=7cm,clip=]{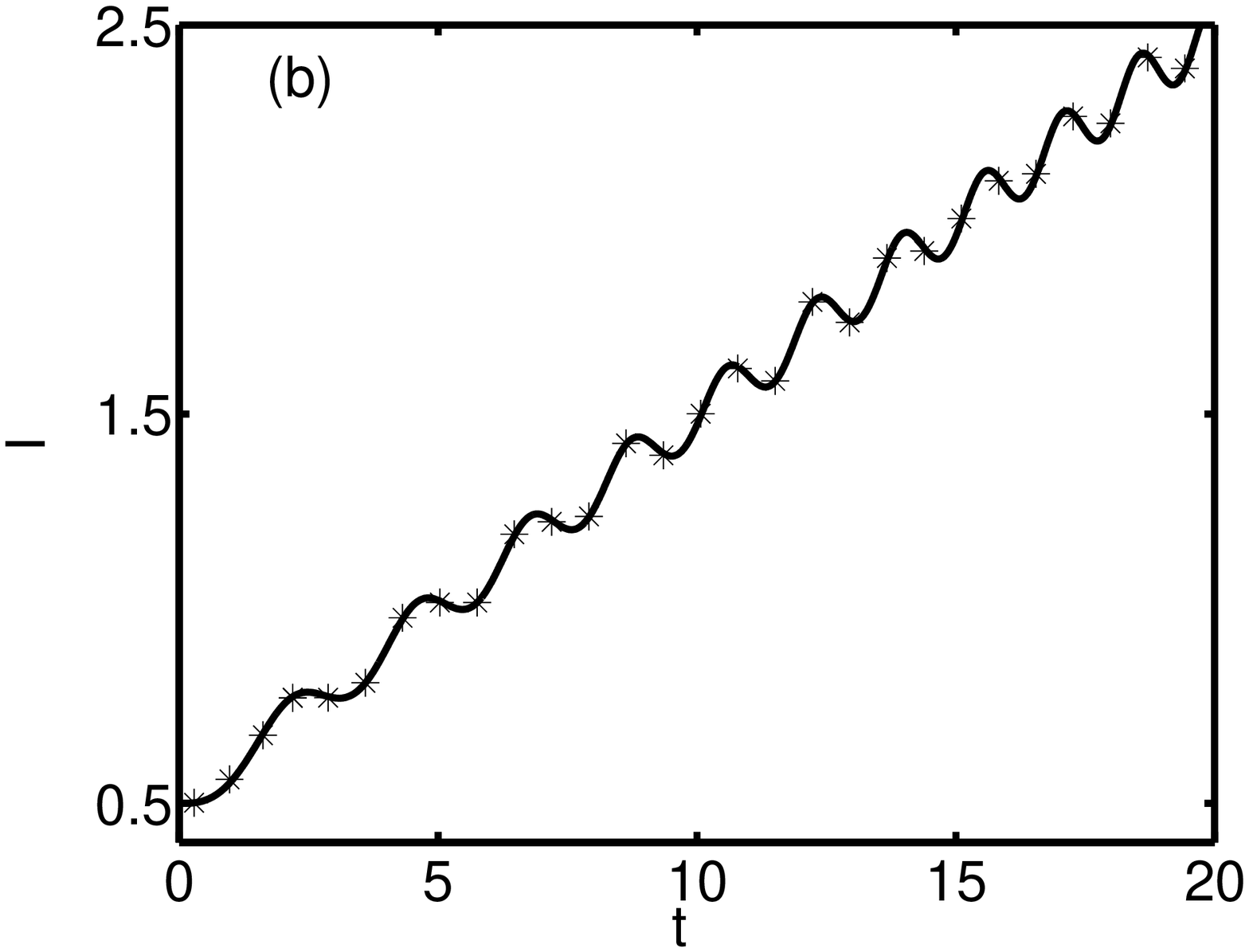}
\caption{The phase $\phi(t)$ (a) and the action $I(t)$ (b) as
functions of time for $m=0.2$, $\phi(0)=0$ and $I(0)=1/2$. The
asterisks mark the asymptotic solutions given by Eqs.
(\ref{solapproxphi}) (a) and (\ref{solI}) (b). The solid lines
mark the numerical solutions.} \label{fitphase}
\end{figure}

Let us linearize Eq. (\ref{diffphi}) in the vicinity of $\phi=0$.
As discussed above, this requires either a small chirp rate, $m
\ll 1$, or a long time, $t \gg 1$. We obtain:
\begin{equation}\label{linphi}
\ddot{\phi}+4(1+mt)\phi=m\,.
\end{equation}
We look for the solution in the form
\begin{equation}\label{phisubs}
\phi(t)=\frac{m}{4(1+m t)}+\delta\phi(t)\,,
\end{equation}
where the first term is the phase trend (\ref{phistar}).
Neglecting higher-order terms, we arrive at the Airy equation
\begin{equation}\label{newlinphi}
\ddot{\delta\phi}+4(1+mt)\delta\phi=0\,,
\end{equation}
whose general solution is \cite{Abramowitz}:
\begin{eqnarray}\label{solprev}
\delta\phi(\tau)&=&C_{1}\,Ai
\left[-\left(\frac{2}{m}\right)^{2/3}(1+
mt)\right]+ \nonumber \\
&&C_{2}\, Bi \left[-\left(\frac{2}{m}\right)^{2/3}(1+
mt)\right]\,.
\end{eqnarray}
Here $Ai(\tau)$ and $Bi(\tau)$ are the Airy functions of the first
and second kind, and $C_{1}$ and $C_{2}$ are constants depending
on the initial conditions. Using the large-argument expansion of
$Ai(-z)$ and $Bi(-z)$ \cite{Abramowitz}, we obtain:
\begin{equation}
\phi(t) \simeq \frac{m}{4(1+m t)}+\frac{A}{(1 + m t)^{1/4}}
\sin\!\left[\frac{4 (1\!+\!mt)^{3/2}}{3 m}+\xi\right]\,,
\label{solapproxphi}
\end{equation}
where $A$ and $\xi$ are new constants. The action $I(t)$ can be
found using the second of Eqs. (\ref{chirp1}):
\begin{equation}
I(t)\simeq\frac{mt+1}{2}-A(1+mt)^{1/4}\cos\!\left[\frac{4
(1\!+\!mt)^{3/2}}{3 m}+\xi\right]\,. \label{solI}
\end{equation}
The first terms in Eqs. (\ref{solapproxphi}) and (\ref{solI}) are
the systematic trends (\ref{phistar}) and (\ref{istar}). The
solutions (\ref{solapproxphi}) and (\ref{solI}) coincide, up to
notation, with the WKB-solutions obtained in Ref. \cite{Khain}.
Fig. \ref{fitphase} shows excellent agreement between the
analytical solutions (\ref{solapproxphi}) and (\ref{solI}) and
numerical solutions.

\subsubsection{Capture into resonance} Now we consider the case
of driving a Faraday wave starting from a very small amplitude and
a large initial detuning, that is far from resonance. Here one is
interested in the capture into resonance. In the case of external
autoresonance this phenomenon has been extensively studied by
Friedland \cite{Friedland1}. In the case of parametric
autoresonance this phenomenon has not been addressed, although
equations similar to (\ref{chirp1}) were analyzed in Ref.
\cite{Lazar7} in the context of the second-harmonic autoresonance
in an externally driven oscillator.

We start, in Eqs. (\ref{chirp}), at a large negative time $t_0<0$,
and assume that the initial detuning  $m t_{0}$ is very large in
the absolute value, while the initial scaled amplitude $B(t=t_0)$
is much less than unity. As long as $B(t)\ll 1$, one can neglect
the $B^{2}$-term in the second of Eqs. (\ref{chirp}):
\begin{equation}\label{ch1}
\dot{\phi} = \cos(2\phi)+ mt\,.
\end{equation}
This equation describes two distinct stages of the dynamics. In
the first stage $t$ is large and negative, and $|mt| \gg 1$.
Therefore, the term $m t$ is dominant, so \cite{Lazar7}
\begin{equation}\label{parabphi}
\phi\simeq\phi(t_0)+\frac{m}{2}(t-t_0)^2\,.
\end{equation}
During this \textit{pre-locking} stage, the phase varies rapidly.
The second stage occurs roughly at $-1/m<t<1/m$. Here the two
terms on the right hand side of Eq. (\ref{ch1}) are comparable. As
$m \ll 1$, the duration of this stage is long compared with unity,
and the phase approaches, on a time scale ${\cal O}(1)$, the
\textit{unstable} quasi-fixed point
\begin{equation}\label{ch2}
\phi_{*}(t)\simeq\frac{\arccos(-mt)}{2}-\frac{m}{4[1-(mt)^{2}]^{1/2}}\,,
\end{equation}
where we have included the next-order correction in $m$. As noted
previously, this quasi-fixed saddle point ceases to exist at
$t>1/m$; the correction term in Eq. (\ref{ch2}) is invalid too
close to $t=1/m$. We call this regime ``linear phase locking"
stage.

\begin{figure}[ht]
\includegraphics[width=7cm,clip=]{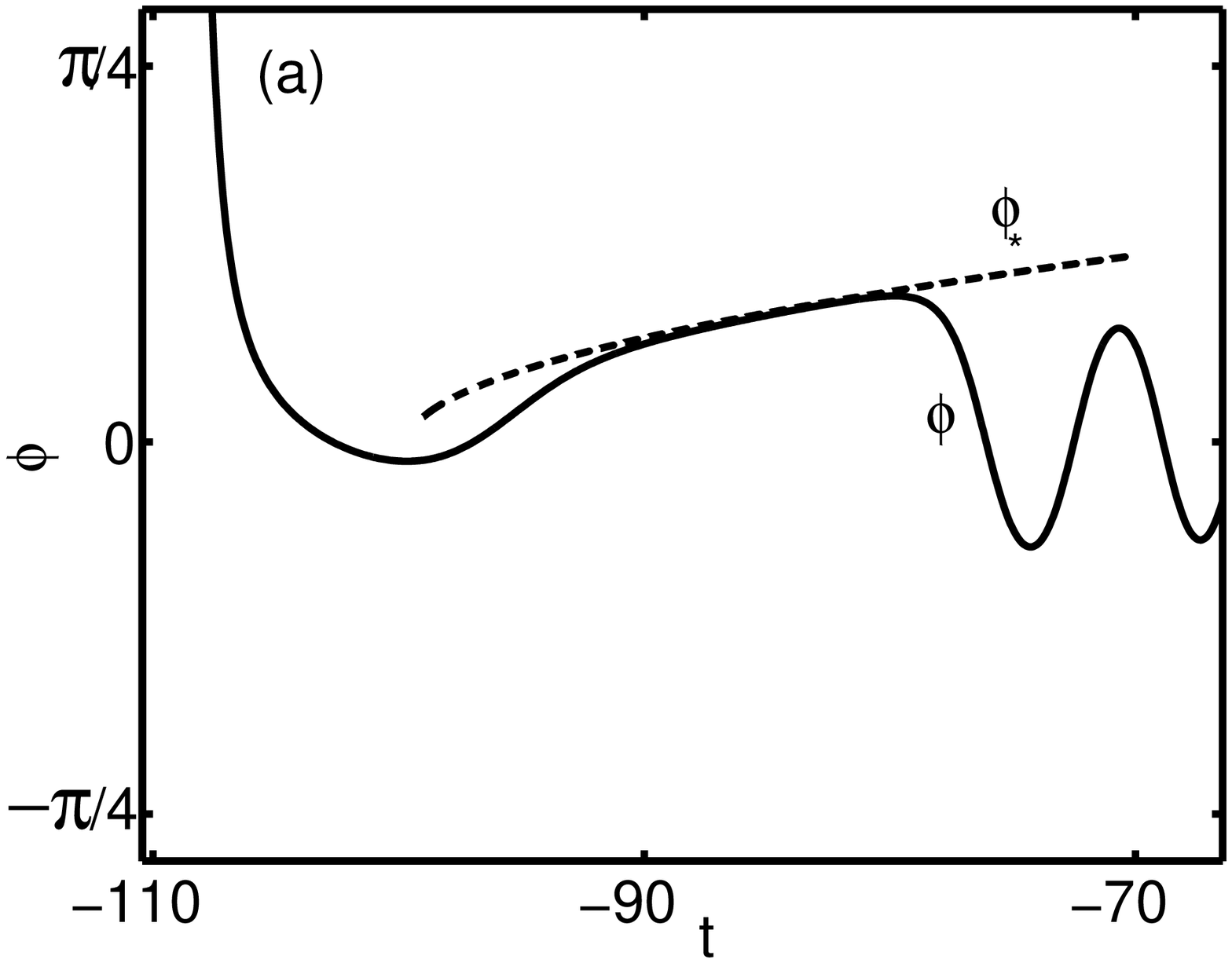}
\includegraphics[width=7cm,clip=]{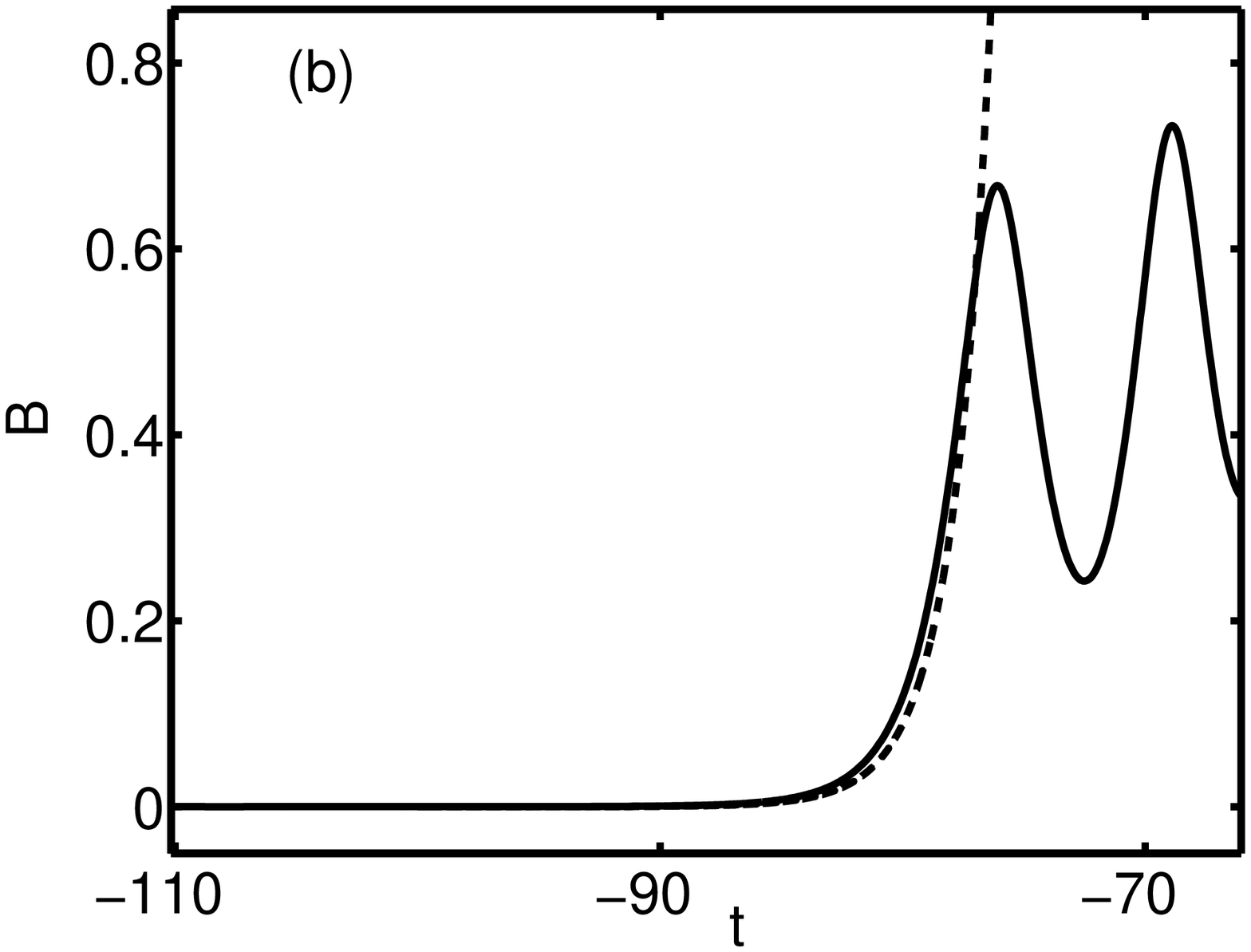}
\caption{The dynamics of the system in the linear phase locking
stage $-1/m<t<1/m$. The phase $\phi(t)$ (a) and amplitude $B(t)$
(b), found numerically, are shown by solid lines. The quasi-fixed
saddle point $\phi_{*}(t)$ [Eq. (\ref{ch2})] (a) and the amplitude
[Eq. (\ref{ampsol2})] (b) are denoted by dashed lines. The initial
conditions are $t_{0}=-400$, $\phi(t_{0})=0$, and
$B(t_{0})=10^{-6}$, the scaled chirp rate $m= 0.01$. It can be
seen that, as $B$ becomes comparable to $1$, the linear theory
breaks down.} \label{ampexp001}
\end{figure}

Let us obtain the dependence of $B$ on time in the pre-locking and
linear phase locking stages. In the pre-locking stage $\phi(t)$ is
given by Eq. (\ref{parabphi}). Then the first of Eqs.
(\ref{chirp}) yields
\begin{equation}\label{ampsol1}
B(t)\simeq B(t_0) \exp\left\{\int_{t_0}^t \sin \left[m (s-t_0)^2+2
\phi(t_0)\right]ds \right\}
\end{equation}
(the integral can be expressed through the Fresnel integral). At
this stage $B(t)$ oscillates rapidly, because of the rapid and
monotonic change of the phase. At the linear phase locking stage,
$\phi(t)\simeq\phi_{*}(t)$ as given by Eq. (\ref{ch2}). Then,
neglecting higher-order terms in $m$, we arrive at the equation
\begin{equation}\label{bsqrt}
\dot{B}\simeq B\sqrt{1-(mt)^{2}}
\end{equation}
which yields
\begin{equation}\label{ampsol2}
B \simeq B_{2}\exp\left\{\frac{1}{2}
\left[t\sqrt{1-(mt)^{2}}+\frac{\arcsin(mt)}{m}\right]\right\}\,,
\end{equation}
where $B_{2}$ is a constant determined by the initial conditions.

Equation (\ref{ampsol2}) breaks down when the earliest of the two
events occurs: $t$ becomes larger than $1/m$, or $B(t)$ becomes
comparable to unity, so that one cannot neglect the $B^2$-term in
the second of Eqs. (\ref{chirp}).  The phase starts to oscillate,
while the amplitude both oscillates and grows, see Fig.
\ref{ampexp001}, and the system enters the autoresonance regime.

Similar results are observed when starting from a small amplitude
and a \textit{zero} initial detuning (that is, in exact linear
resonance). Therefore, when the initial amplitude is very small,
and $m\ll\,1$, phase locking is very robust.

\subsubsection{Critical chirp rate and the limiting trajectory}

As we have seen numerically, when $m$ approaches the critical
value $m_{cr}$, the phase $\phi$ approaches $\pi/2$ at $t \gg 1$.
Let us assume that, at a time $t_{0}$, $\phi$ is already in the
vicinity of $\pi/2$: $\phi=\pi/2-\delta \phi$, where $\delta \phi
\ll 1$. Linearizing Eq. (\ref{diffphi}), we obtain
\begin{equation}\label{linphipi}
\ddot{\delta\phi}-4\delta\phi(mt-1)=-m\,.
\end{equation}
To find the trend, we neglect the term $\ddot{\delta\!\phi}$.
Therefore, at $t\gg 1$, we obtain $\delta \phi \simeq 1/(4 t)$, so
\begin{equation}
\phi_*(t)\simeq\frac{\pi}{2}-\frac{1}{4t}\,. \label{1/4t}
\end{equation}
Using the second of Eqs. (\ref{chirp1}), we obtain the respective
trend of $I(t)$:
\begin{equation}\label{appI}
I_*(t)\simeq\frac{m t}{2}-\frac{1}{2}\,.
\end{equation}
At this stage we notice that Eqs. (\ref{1/4t}) and (\ref{appI})
describe, at $t \gg 1$, one of the unstable (saddle) quasi-fixed
points of the system: the one that appears close to $t=1/m$. This
again shows that adiabaticity holds not only at $m \ll 1$, but
also at $t \gg 1$.

\begin{figure}[ht]
\includegraphics[width=7cm,clip=]{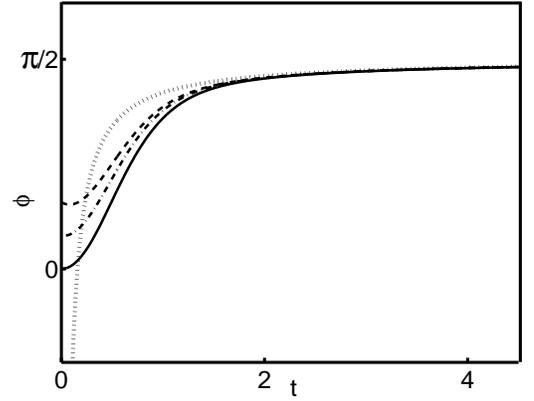}
\caption{Numerical solutions for the phase $\phi(t)$, for three
different initial conditions, versus the universal asymptotics
(\ref{1/4t}), denoted by the dotted line. The solid, dashed-dotted
and dashed lines show the phase dynamics when starting from
$\phi(0)=0$ (here $m \simeq m_{cr}\simeq\,4.963$), $1/4$ (here $m
\simeq m_{cr}\simeq\,6.237$), and $1/2$ (here $m \simeq
m_{cr}\simeq\,7.094$), respectively.  $I(0)=1/2$ in all three
cases.} \label{analcompnumfix}
\end{figure}

\begin{figure}[ht]
\includegraphics[width=7cm,clip=]{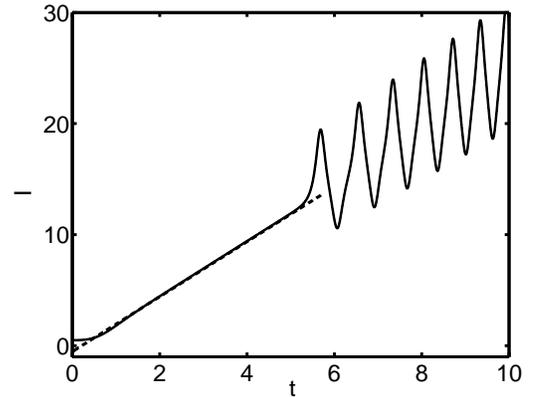}
\caption{Shown is $I(t)$ for $m$ just below $m_{cr} \simeq 4.963$.
The solid line shows the numerical solution, starting from
$I(0)=1/2$ and $\phi(0)=0$. The dashed line shows the trend
(\ref{appI}).}\label{analcompnum}
\end{figure}

Now consider small deviations from the unstable trends
(\ref{1/4t}) and (\ref{appI}). Putting
$\phi=\pi/2-1/(4t)-\delta\phi(t)$, we again arrive at the Airy
equation for $\delta\phi(t)$. Its general solution is
$\delta\phi(t)=C_{1}Ai[(4m)^{1/3}t]+C_{2}Bi[(4m)^{1/3}t]$, where
$C_{1}$ and $C_{2}$ are determined by the initial conditions.
Assuming $m^{1/3} t \gg 1$, we employ the asymptotics of the Airy
functions \cite{Abramowitz}:
\begin{equation}\label{limitai}
Ai(z)\simeq\frac{\exp\left[-(2/3)z^{3/2}\right]}{2\sqrt{\pi}z^{1/4}}\,,\;\;Bi(z)\simeq\frac{\exp\left[(2/3)z^{3/2}\right]}{\sqrt{\pi}z^{1/4}}
\end{equation}
at $z\gg\,1$. The exponentially decaying term is negligible. The
exponentially growing term describes instability of the special
trajectory with respect to small perturbations. This instability
occurs both at $m<m_{cr}$ (when the phase leaves the vicinities of
the saddle point and goes to the vicinity of the stable
quasi-fixed point), and at $m>m_{cr}$ (when the phase locking
terminates). For the special trajectory, obtained for $m=m_{cr}$,
the coefficient $C_{2}$ must vanish, which brings us back to Eqs.
(\ref{1/4t}) and (\ref{appI}), where we must put $m=m_{cr}$. Now
it is clear that the asymptotic behavior of the $\phi_*(t)$ at
$m=m_{cr}$  is independent of $m_{cr}$ and, therefore, on the
initial conditions. On the contrary, the asymptotic behavior of
$I_*(t)$ at $m=m_{cr}$ does depend on $m_{cr}$ and, therefore, on
the initial conditions. Unfortunately, the local analysis does not
enable one to find the value of $m_{cr}$ analytically.

Figures \ref{analcompnumfix} and \ref{analcompnum} show the
behavior of the system when $m$ is very close to (just below)
$m_{cr}$. Figure \ref{analcompnumfix} shows that, after a time
${\cal O}(1)$, the phase follows the trend (\ref{1/4t}),
independently of the initial conditions (and of the value of
$m_{cr}$).  Figure \ref{analcompnum} shows that, after a time of
${\cal O}(1)$, good agreement between $I(t)$, found numerically,
and the trend (\ref{appI}) holds until the time when the
``particle" leaves the vicinity of the unstable point and
transfers to the vicinity of the stable point.

\section{Chirped Faraday waves in a low-viscosity fluid}

\begin{figure}
\includegraphics[width=7cm,clip=]{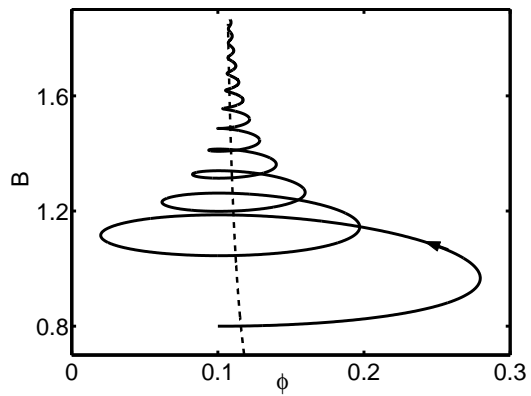}
\caption{Parametric autoresonance in the presence of viscosity.
Shown is a projection of a three-dimensional trajectory in the
space of $\phi$, $B$ and $t$, upon the $(\phi, B)$-plane.  The
solid line shows a numerical solution, the dashed line shows the
trends (\ref{mainvisc1}). After several nonlinear cycles, the
trajectory converges into a (slowly time-dependent) stable focus,
which moves upwards. The parameters are $\Gamma=0.2$, $m=0.1$,
$B(0)=0.8$, $\phi(0)=0.1$, and a zero initial detuning.}
\label{phasediagchirpvisc}
\end{figure}

We now account for a small viscosity and briefly describe the
dynamics of the phase and amplitude of the primary mode in the
case of a slowly time-dependent vibration frequency. The weakly
nonlinear governing equations are obtained by analogy with Eqs.
(\ref{mainvisc}):
\begin{eqnarray}
\dot{B}&=&B\sin(2\phi)-\Gamma\,B\,,\nonumber\\
\dot{\phi}&=&\cos(2\phi)-B^{2}+mt\,, \label{mainvisc1}
\end{eqnarray}
where $t$ is the slow time as before. For $\Gamma< 1$, there
exists a stable quasi-fixed point which describes autoresonant
excitation of the wave:
\begin{eqnarray}
B_{*}&\simeq&\left[(1-\Gamma^{2})^{1/2}+mt\right]^{1/2}\,,\nonumber\\
\phi_{*}&\simeq&\frac{1}{2}\arcsin(\Gamma)+\frac{m}{4[1-\Gamma^{2}+mt(1-\Gamma^{2})^{1/2}]}\,.
\label{bphistarfull}
\end{eqnarray}
When $\Gamma\ll 1$,  Eqs. (\ref{bphistarfull}) become
\begin{equation}\label{bphistar}
B_{*}\simeq\sqrt{1+mt}\,,\;\;\;\phi_{*}\simeq\frac{\Gamma}{2}+\frac{m}{4(1+mt)}\,.
\end{equation}
Figure \ref{phasediagchirpvisc} shows a projection on the
$(\phi,B)$ plane of a three-dimensional trajectory in the space of
$\phi$, $B$ and $t$. One can see phase locking and a steady growth
of the wave amplitude with time.

\begin{figure}[ht]
\includegraphics[width=7cm,clip=]{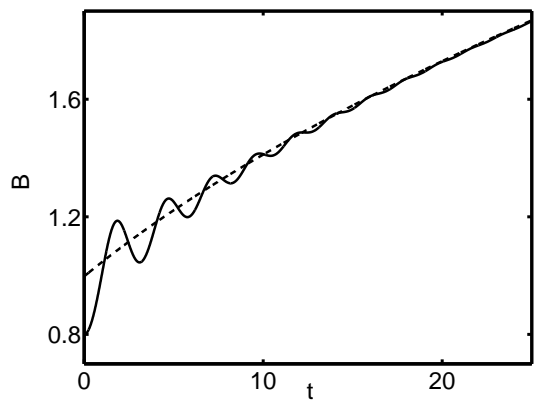}
\caption{Parametric autoresonance in the presence of viscosity.
Shown is  the amplitude $B(t)$ versus time for a subcritical
damping. The solid line denotes a numerical solution, while the
dashed line shows the trend $B_{*}(t)$, see Eq.
(\ref{bphistarfull}). The parameters are the same as in Fig.
\ref{phasediagchirpvisc}.} \label{chirptrendvisc}
\end{figure}

\begin{figure}
\includegraphics[width=7cm,clip=]{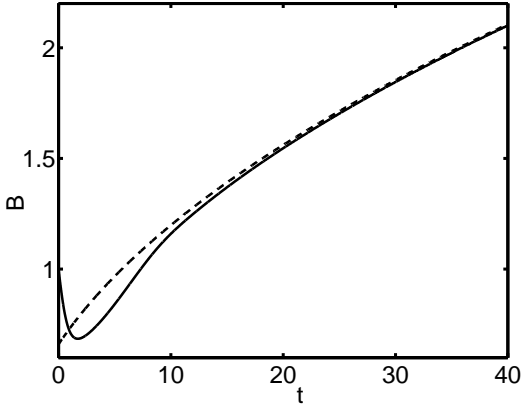}
\vspace{0.3cm} \caption{Parametric autoresonance in the presence
of viscosity. Shown is the amplitude $B(t)$ versus time for a
supercritical damping. The solid line denotes the amplitude $B(t)$
found numerically, and the dashed line denotes its trend
$B_{*}(t)$, see Eq. (\ref{bphistarfull}). The parameters are
$\Gamma=0.9$, $m=0.1$, $B(0)=1$, $\phi(0)=0.25$, and a zero
initial detuning. } \label{chirptrendstrongvisc}
\end{figure}

\begin{figure}[ht]
\includegraphics[width=7cm,clip=]{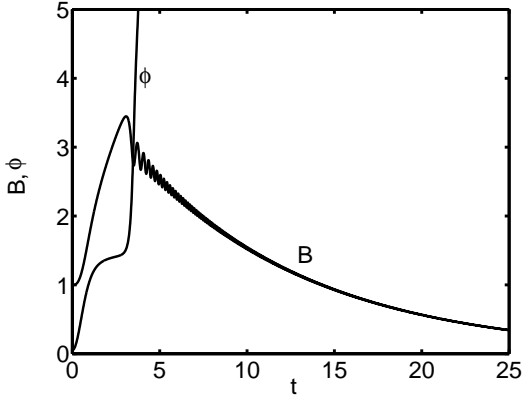}
\caption{Breakdown of autoresonance in the presence of viscosity.
Shown are the amplitude $B(t)$ and the phase $\phi(t)$, for
$m\simeq\,4.4$ (just above $m_{cr}$) for the initial conditions
$B(0)=1$ and $\phi(0)=0.05$, and for $\Gamma=0.1$. One can see
that, after a strong transient excitation, the phase unlocks, and
the amplitude decays.} \label{decayamp}
\end{figure}

Figures \ref{chirptrendvisc} and \ref{chirptrendstrongvisc} show
two different regimes of autoresonant growth in the dissipative
system.  Shown in the two figures are the primary mode amplitude
versus time for the scaled damping rates $\Gamma<\Gamma_{cr}$ and
$\Gamma>\Gamma_{cr}$, respectively. As the initial detuning is
zero, $\Gamma_{cr}=2/\sqrt{5} = 0.89\dots$ here, see Eq.
(\ref{gammacrit}). Figure \ref{chirptrendvisc} shows decaying
oscillations on top of the amplitude growth, given by the first of
Eqs. (\ref{bphistarfull}).  Figure \ref{chirptrendstrongvisc}
shows a non-oscillatory regime of the amplitude growth. Figure
\ref{decayamp} shows breakdown of autoresonance when the chirp
rate $m$ exceeds a critical value.


\begin{figure}[ht]
\includegraphics[width=7cm,clip=]{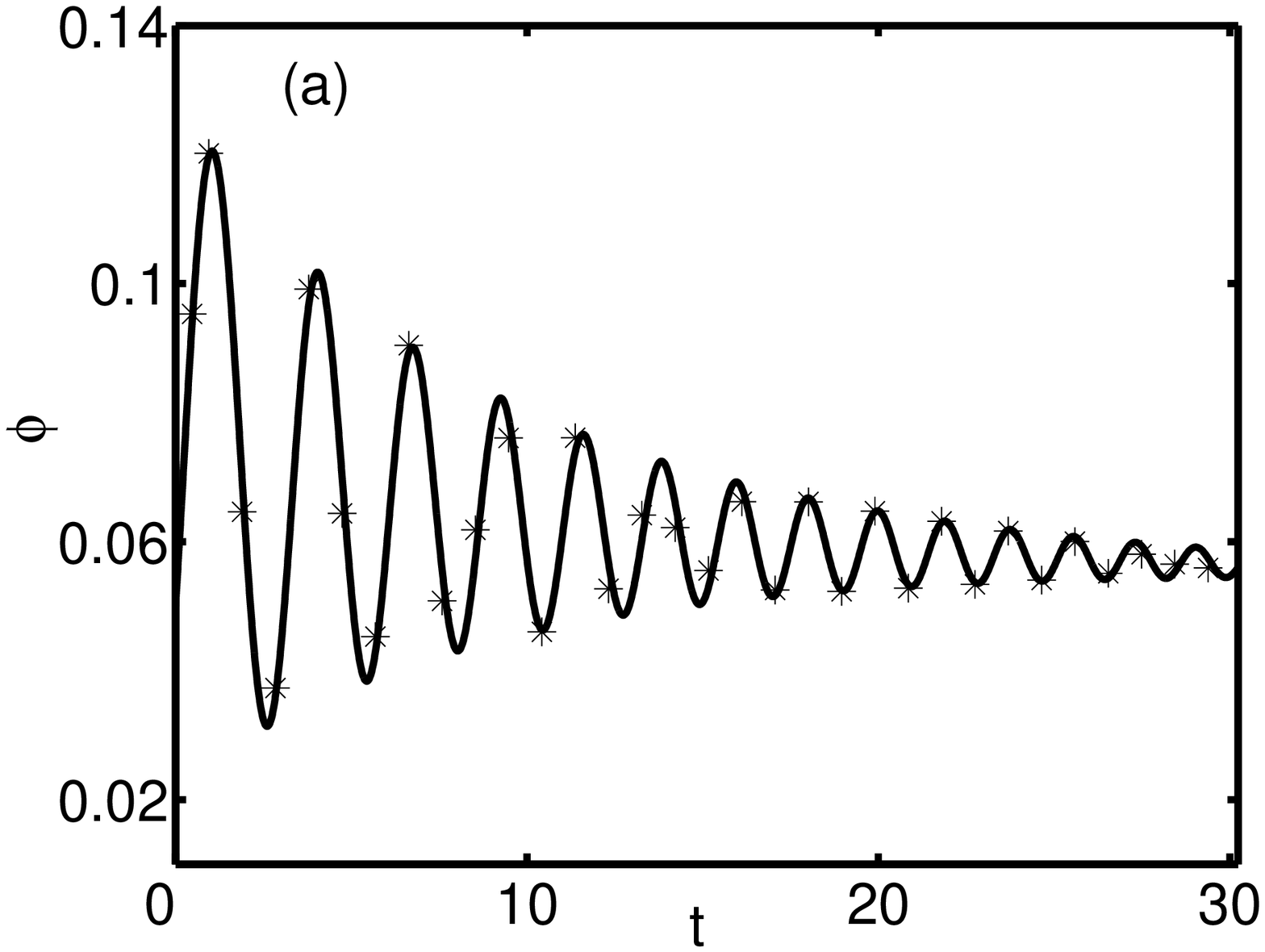}
\includegraphics[width=7cm,clip=]{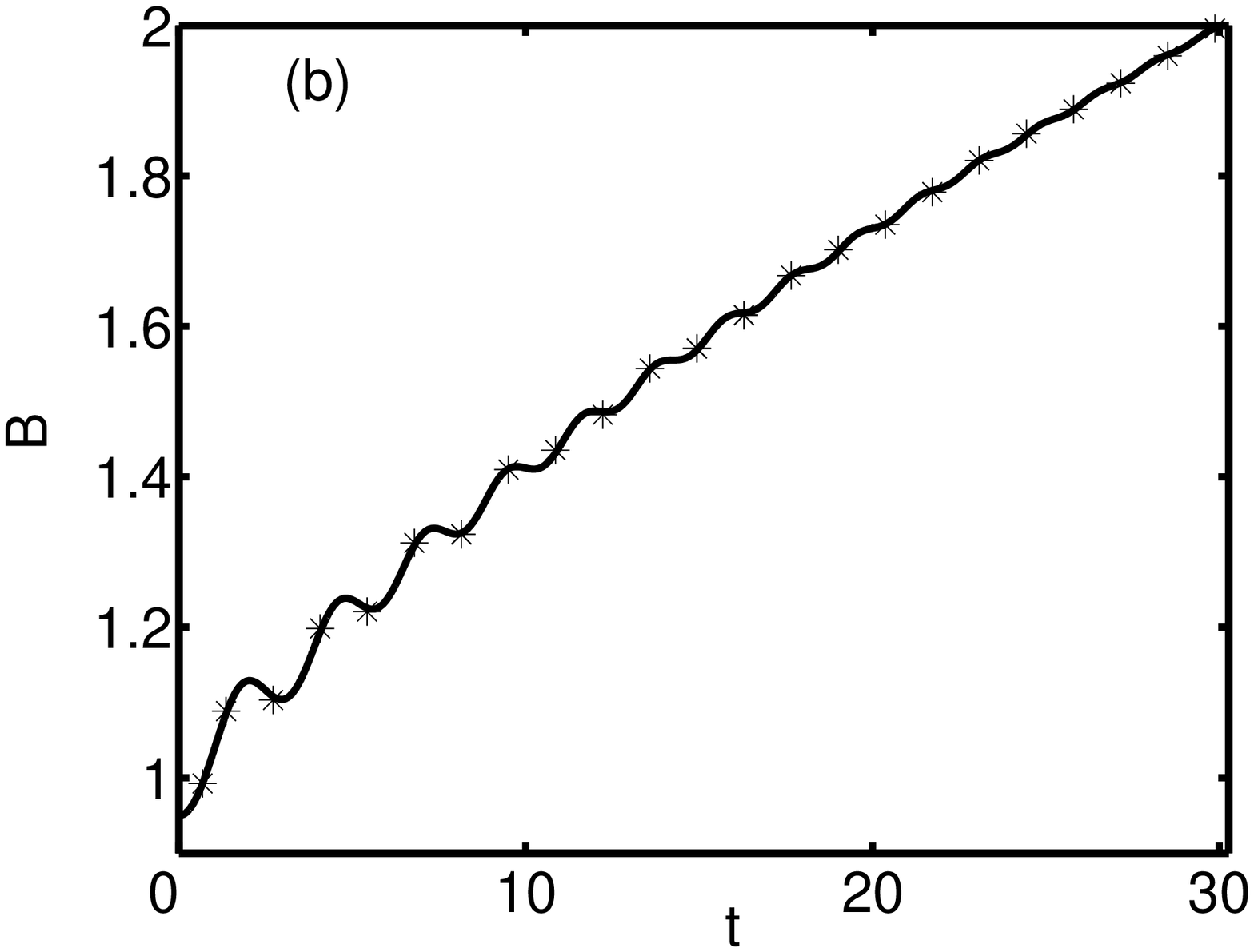}
\caption{The phase $\phi(t)$ (a) and the amplitude $B(t)$ (b) as
functions of time in the presence of small viscosity. The
parameters are $m=0.1$, $\Gamma=0.1$. The initial conditions are
$\phi(0)=0.05$ and $B(0)=0.95$. The asterisks mark the analytical
solutions, the solid lines show numerical solutions.}
\label{fitampvisc}
\end{figure}

Similarly to the inviscid case, one can rewrite the governing
equations (\ref{mainvisc1}) as a single equation for the complex
amplitude, $\psi = B \exp (i\phi)$:
\begin{equation}\label{NLSvisc}
i\psi_{t}+\psi^{*}-(|\psi|^{2}-mt-i\Gamma)\psi=0\,.
\end{equation}
Assuming a solution growing in time (that is, phase locked
solution), we obtain, at $t \gg 1$, $B=|\psi|\simeq (mt)^{1/2}$,
as before.

Equations (\ref{mainvisc1}) can be also rewritten as a single
second-order equation for $\phi(t)$:
\begin{eqnarray}\label{diffphivisc}
\ddot{\phi}+2 \Gamma
\dot{\phi}&+&\sin(4\phi)+2mt\sin(2\phi) \nonumber\\
&-&2\Gamma \cos(2\phi)= 2 \Gamma m t+m
\end{eqnarray}
[compare it with Eq. (\ref{diffphi})]. This equation is convenient
for a perturbative treatment in the vicinity of the stable
quasi-fixed point.  For $\Gamma \ll 1$, we can linearize Eq.
(\ref{diffphivisc}) around $\phi=0$:
\begin{equation}\label{linphivisc}
\ddot{\phi}+2\Gamma\dot{\phi}+4 (1+mt)\phi = 2\Gamma\!(1+mt)+m\,.
\end{equation}
Now we substitute $\phi(t)=\phi_*(t)+\delta\phi(t)$, where
$\phi_*$ is given by the second of Eqs. (\ref{bphistar}), and
obtain a linear equation
\begin{equation}
\ddot{\delta\phi} + 2\Gamma\dot{\delta\phi} + 4(1+ m t) \phi =
0\,.
\end{equation}
Its approximate solution, at $m \ll  1$ and $\Gamma \ll 1$, can be
written as
\begin{equation}\label{phidiss}
\delta\phi \simeq \frac{A \, e^{-\Gamma t}}{(1+mt)^{1/4}} \sin
\left[\frac{4 (1+ m t)^{3/2}}{3 m} + \xi\right]\,,
\end{equation}
where $A$ and $ \xi$ are constants depending on the initial
conditions. The respective solution for $B(t)$ is
\begin{equation}\label{Bdiss}
B \simeq (1+mt)^{1/2}-\frac{A\,e^{-\Gamma t}}{(1+mt)^{1/4}} \cos
\left[\frac{4 (1+ m t)^{3/2}}{3 m} + \xi\right]\,,
\end{equation}
These solutions are simple extensions of the non-viscous
perturbative solutions. A comparison with numerical solutions is
shown in Fig. \ref{fitampvisc}, and excellent agreement is
observed.

\begin{figure}[ht]
\includegraphics[width=7cm,clip=]{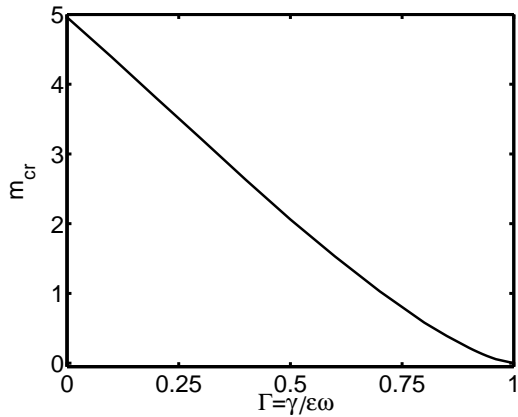}
\caption{The critical scaled chirp rate $m_{cr}=\mu_{cr}/(\varepsilon^2 \omega_1^2)$ versus the scaled damping
rate of the wave $\Gamma=\gamma/(\varepsilon \omega_1)$, as described by Eqs. (\ref{mainvisc1}).  The initial
amplitude and phase correspond to the stable fixed point of Eqs. (\ref{mainvisc}) with $\Delta=0$.}
\label{critmvisc}
\end{figure}

In the viscous case the critical value $m_{cr}$ of the scaled
chirp rate $m=\mu/(\varepsilon \omega_1)^{2}$ depends on the
scaled damping rate of the wave $\Gamma=\gamma/(\varepsilon
\omega_1)$. Figure \ref{critmvisc} shows this dependence, which we
found numerically when starting at $t=0$ from the stable fixed
point of Eqs. (\ref{mainvisc}) with $\Delta=0$, that is from
$B_{*}=(1-\Gamma^2)^{1/4}$ and $\phi_{*}=(1/2) \arcsin (\Gamma)$.
One can see that, as $\Gamma$ increases, the critical chirp rate
goes down monotonically. As $\Gamma$ approaches $1$, $m_{cr}$ goes
to $0$. Notice that, once we return to the physical (dimensional)
critical chirp rate $\mu_{cr}$ and the wave damping rate $\gamma$,
the dependence of $\mu_{cr}$ on $\varepsilon$, at fixed $\gamma$,
is \textit{not} a power law.

\begin{figure}[ht]
\includegraphics[width=7cm,clip=]{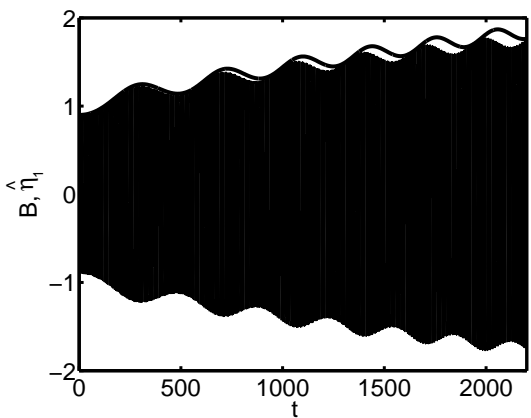}
\caption{$B(t)$ (the thick solid line) and $\hat{\eta}_{1}$ (the
thin solid line) versus time. The parameters are $\Gamma=0.006$,
$m=0.18$, $\varepsilon=0.006$, $\hat{\eta}_{1}(0)=B(0)=0.9$ and
$\dot{\hat{\eta}}_{1}(0)=\phi(0)=0$. } \label{fullnumerical}
\end{figure}

Finally, we tested the accuracy of our reduced equations
(\ref{mainvisc1}). We compared numerical solutions of Eqs.
(\ref{mainvisc1}) with numerical solutions of the unreduced
equation of motion (\ref{motionchirp}) for the primary mode
amplitude $\eta_{1}(t)$, with a linear damping term added.
Rescaling the time $\tau=\omega_{1}t$, and the amplitude
$\hat{\eta}_{1}=(k_{1}\eta_{1})/(2\sqrt{\varepsilon})$, one can
rewrite the unreduced equation as
\begin{eqnarray}
\ddot{\hat{\eta}}_{1}&+&2\varepsilon\Gamma\dot{\hat{\eta}}_{1}+10\varepsilon\dot{\hat{\eta}}_{1}^{2}\hat{\eta}_{1}-6\varepsilon\hat{\eta}_{1}^{3}+\nonumber\\
&+&\hat{\eta}_{1}\left[1+4\varepsilon\cos
\left(2\tau-m\varepsilon^{2}\tau^2\right)\right]=0\,.
\label{fullunreduced}
\end{eqnarray}
A typical example of this comparison is shown in Fig.
\ref{fullnumerical}, and a fairly good agreement between the
envelope of $\hat{\eta}_{1}(t)$ and the amplitude $B(t)$ is
observed.

\section{Discussion}

This paper presents a theory of weakly nonlinear standing gravity
waves, parametrically excited by weak vertical vibrations with a
down-chirped vibration frequency. We have shown that autoresonance
phase locking and a steadily growing wave amplitude can be
achieved despite the nonlinear frequency shift of the wave. For
typical initial conditions we have found the critical chirp rate,
above which autoresonance breaks down. When starting from a very
small wave amplitude, and slowly passing through resonance, phase
locking always occurs. We have obtained approximate analytical
expressions for the time-dependent wave profile in different
regimes. We have demonstrated that each of the \textit{three}
quasi-fixed points of the reduced dynamic equation, describing the
primary mode, plays an important role in the dynamics of the
system and/or in determining the critical chirp rate.

Parametric autoresonance in Faraday waves is a robust phenomenon,
and its experimental observation should not be difficult. To test
our theory, one should use a quasi-two-dimensional tank with a
low-viscosity liquid, and perform measurements of the standing
wave elevation as a function of time. The long-term wave amplitude
trend [see the first of Eqs. (\ref{bphistarfull})], and the
critical chirp rate (see Fig. 20), give examples of quantitative
predictions of the theory that can be tested in experiment. One
such experiment is presently under way \cite{Oded}.

The applicability of the weakly nonlinear theory presented in this work is limited to \textit{weak} forcing  and
\textit{weak} damping. Our analysis neglected higher order terms in $\varepsilon$, which include a nonlinear
forcing \cite{Miles3}, a cubic damping \cite{Miles3,douady,milner}, a small correction to the linear
detuning/frequency shift, a quintic conservative term \cite{decent}, etc. These terms can be included in the
theory of autoresonance in order to achieve a better accuracy. Importantly, once these higher-order terms are
added to the constant-frequency model [Eq. (\ref{mainvisc})], the non-trivial stable fixed point will exist only
up to a certain value of the frequency detuning \cite{Miles3,douady,milner,decent}. Therefore, in experiment,
the autoresonant growth of the wave is expected to terminate when the (time-dependent) frequency detuning comes
close to the maximum value, for which the non-trivial stable fixed point in the underlying constant-frequency
model still exists. The breakdown of autoresonance is expected to occur at an amplitude smaller than the
wave-breaking amplitude, which is comparable to the wavelength
 \cite{Schultz}.

\section*{ACKNOWLEDGEMENTS}

We thank Oded Ben-David and Jay Fineberg for useful discussions.
This research was supported by the Israel Science Foundation.


\begin{references}
\bibitem {constamp} We checked that, when the vibration amplitude $a_{0}$ is kept
constant (so that $\varepsilon$ varies with time), qualitatively
similar autoresonance excitation regime is observed, provided the
chirp rate is sufficiently small.
\end{references}
\end{document}